\documentclass[aps,pra,twocolumn,groupedaddress,showpacs,amsmath,amssymb]{revtex4-1}

\usepackage{graphicx, comment}
\usepackage{dcolumn}
\usepackage{bm}

\begin{document}


\title{Entangled plasmon generation in nonlinear spaser system under action of external magnetic field}


\author{M.Yu. Gubin}
\affiliation{Department of Physics and Applied Mathematics,
Vladimir State University named after A. G. and N. G. Stoletovs, Gorky str. 87, Vladimir,
600000, Russia}
\author{A.V. Shesterikov}
\affiliation{Department of Physics and Applied Mathematics,
Vladimir State University named after A. G. and N. G. Stoletovs, Gorky str. 87, Vladimir,
600000, Russia}
\author{S.N. Karpov}
\affiliation{Department of Physics and Applied Mathematics,
Vladimir State University named after A. G. and N. G. Stoletovs, Gorky str. 87, Vladimir,
600000, Russia}
\author{A.V. Prokhorov}
\email{avprokhorov33@mail.ru}
\affiliation{Department of Physics and
Applied Mathematics, Vladimir State University named after A. G. and N. G. Stoletovs, Gorky str. 87, Vladimir,
600000, Russia}


\date{\today}

\begin{abstract}
The present paper theoretically investigates features of quantum dynamics for localized plasmons in three-particle or four-particle spaser systems consisting of metal nanoparticles and semiconductor quantum dots. In the framework of the mean field approximation, the conditions for the observation of stable stationary regimes for single-particle plasmons in spaser systems are revealed, and realization of these regimes is discussed. The strong dipole-dipole interaction between adjacent nanoparticles for the four-particle spaser system is investigated. We show that this interaction can lead to the decreasing of the autocorrelation function values for plasmons. The generation of entangled plasmons in a three-particle spaser system with nonlinear plasmon-exciton interaction is predicted. For the first time the use of an external magnetic field is proposed for control of the cross-correlation between plasmons in the three-particle spaser system.
\end{abstract}


\maketitle

\section{\label{Int}Introduction}
Recent rapid progress in the field of nanotechnology has led to the practical possibility of the generation and control of N-photon states using single quantum emitters~\cite{1} and high-quality micro-~\cite{2} and nanoresonators~\cite{3}. The creation of the nanolaser is key to the development of such devices~\cite{4}. The functional features of the nanolaser require a reformulation of the known rules of laser generation for subwavelength scales~\cite{5}. This can be done on the basis of the model of the localized spaser~\cite{6}. In simple form, this device consists of a semiconductor quantum dot (QD) and a metal nanoparticle (NP) coupled by near-field interaction. The QD is used here as a powerful near-field pump source, since the excitonic decay in the QD leads to a strong perturbation of electronic density in the NP. This perturbation leads to the generation of the plasmons localized on the NP surface~\cite{2ref,7,8}. Reducing the distance between the NP and QD results in an increase of non-radiative energy exchange in the system. If the Rabi frequency of interaction (for example, a dipole-dipole interaction) between nano-objects becomes greater than the characteristic times of decay, then a strong coupling arises in the system~\cite{3ref}.
A realistic model of the spaser can be based on a compound nanoobject consisting of a metal core and a semiconductor shell~\cite{9} or on a distributed system of nanoparticles with complex geometry~\cite{10,10a,11,11a,12}. Currently, both localized subwavelength~\cite{4} and waveguide~\cite{13} spaser-like systems are implemented.

Chains of near-field coupled NPs and QDs are of significant interest as a platform for quantum computing~\cite{14,14a,14b} with single-photon and single-plasmon states~\cite{15}. For example, the appearance of bunching (antibunching) effects for emitted photons can be observed in self-assembled QD structures~\cite{19} or in the mesoscopic chromophore ensemble strongly coupled with a plasmonic cavity~\cite{16a}. In another approach, a pair of coupled QDs can be used as a powerful source of entangled photons due to the correlations between QD excitons in biexcitonic states~\cite{16}. On the other hand, if QDs are coupled with shared NP~\cite{17}, direct transmission of quantum correlations from bound excitons to plasmons can be achieved. One of the main advantages of such correlated plasmons is the possibility of simpler external control and manipulation of their carriers -- NPs. In particular, this addressing can be carried out using the epifluorescence microscopy technique for single quantum objects~\cite{18}. However, a very important question arises: whether the nonclassical plasmon states will survive if a strong dipole-dipole interaction between NPs is realized.

The purpose of this paper is to optimize the chemical and geometric characteristics of multiparticle dissipative spaser systems in order to achieve a balance of gains and losses for plasmons in accordance with the principles of PT-symmetry~\cite{4ref}. Starting from this paradigm, we consider the possibility of control over the quantum statistical and correlation properties of plasmons generated in such a spaser system.

This paper is structured as follows. In Section~\ref{sec2} we present the model of a $2\times 2$ spaser system as an extension of the model suggested in~\cite{17}. Our model spaser system consists of two
closely located identical NPs with plasmon resonance frequency $\omega_{p}$ and two two-level QDs with exciton energy $\hbar\omega$. All particles in this system are coupled by dipole-dipole interactions. Meanwhile, provided the nanoobjects are located in the vertices of a square, the efficiency of dipole-dipole interactions between the NPs significantly exceeds the efficiency of the interactions between the QDs and for QD-NP pairs. We found stable stationary solutions for this system and investigated the quantum statistics of plasmons localized on the NPs. We have shown that strong interaction between NPs can significantly change the statistical properties of localized plasmons.

In Section~\ref{sec3} we develop a novel model including a configuration of three nanoobjects (NP-QD-NP) coupled by nonlinear dipole-dipole interactions in the presence of an external magnetic field. The nonlinear regime of this ensemble corresponds to the two-quantum processes of the QD biexciton decay in the case $\left|\delta\right| > \Omega_{1,2},\left|\Delta\right|$, where $\delta=\bar{\omega}-\omega$ and $\Delta=\bar{\omega}-\omega_{p}$, $\bar{\omega}$ is the spasing frequency, and $\Omega_{1,2}$ are the Rabi frequencies of dipole-dipole interactions between the QD and NPs. As a result of this nonlinear process, one can expect the appearance of strongly correlated plasmon pairs. The best regime will be one in which each plasmon from such a pair can be localized on its own NP. We focus on this regime because it is useful for the generation of a nonclassical N-particle~\cite{21} and entangled plasmon states~\cite{20} as in quantum optics~\cite{23,24,24a}. The principal difference of our approach from previous studies~\cite{17} is the presence of an external magnetic field, which leads to a change in the QD energy levels and provides a means to control the quantum properties of generated plasmons.

In the technical framework, the presented spaser systems can be experimentally implemented on the basis of nanoparticles assembled on a template patterned in a thin photoresist film~\cite{1ref,161from1ref1}. Such systems can be integrated in the individual plasmonic waveguides~\cite{26} and plasmonic circuits for quantum information processing~\cite{27,27a,27b}.

\section{\label{sec2}The formation of non-classical states of plasmons in the system of two spasers coupled by strong dipole-dipole interactions}
\begin{figure}[t]
\includegraphics[width=\columnwidth]{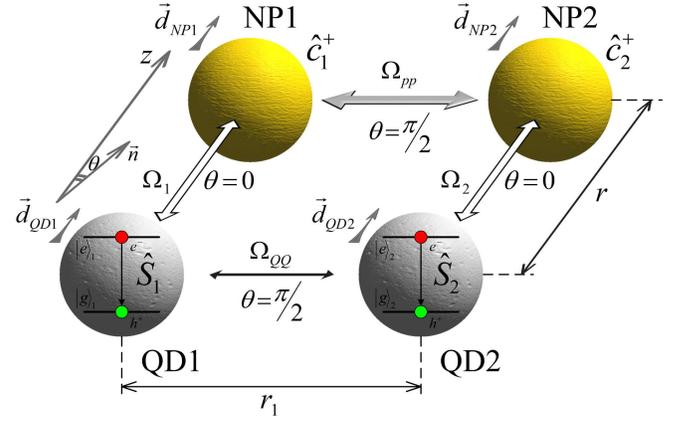}
\caption{\label{fig:1}A four-particle $2\times 2$ spaser model, consisting of two NPs and two QDs.}
\end{figure}

Let us consider the system of two spasers, consisting of 2 QDs and 2 NPs (so-called spaser $2\times 2$), see Fig.~\ref{fig:1}. First of all, the efficiency of interaction in the system will depend on the geometry of the system, where the characteristic lengths are  $r_{NP} $, which is the distance between adjacent NPs, $r_{QD} $, which is the distance between adjacent QDs, and $r_{QN} $, which is the distance in a QD-NP pair. The vector $\vec{n}=\vec{r}/r$, making the angle $\theta $ with axis $\vec{z}$, determines the direction between the centers of any two interacting particles. This spaser system can be considered a part of a more complex hybrid nanostructure that supports the possibility of external effective nonradiative pumping of the system~\cite{7ref}. The assembly of individual nano-objects in the spaser system can be realized by atomic force microscopy~\cite{8ref}. We assume that all dipole moments $\hat{d}_{QD} $ of QD and $\hat{d}_{NP} $ of NP are collinear to each other and parallel to axis $\vec{z}$~\cite{6}. In this paper we do not consider higher orders of multipoles, although intermode interactions for this model can lead to the manifestation of the well-known Fano resonance effect~\cite{5ref}.

The field $\hat{{\rm E} }_{NPi} =-\nabla \hat{A}_{NPi} $ at a distance of $r$ from the $i$--th NP of spherical shape with radius $a_{NPi} $ can be expressed through the vector potential operator $\hat{A}_{NPi} =\mathop{\sum }\limits_{n} \left(\frac{a_{NPi} }{r} \right)^{n+1} Y_{nm} \left(\theta ,\varphi \right)E_{nm} \left(\hat{c}_{i} +\hat{c}_{i}^{+} \right)\vec{e}_{NPi} $~\cite{7}, where $\hat{c}_{i} $($\hat{c}_{i}^{+} $) are annihilation (creation) operators of the plasmon mode in quasistatic approximation, $Y_{nm} \left(\theta ,\varphi \right)=\sqrt{\frac{2n+1}{4\pi } \frac{\left(n-m\right)!}{\left(n+m\right)!} } P_{n}^{m} \left(\cos \theta \right)e^{im\varphi } $ are spherical functions expressed via Legendre polynomials, $\vec{e}_{NP} $ defines the orientation of the NP dipole moment, $E_{nm} =\sqrt{\frac{\hbar \omega _{nm} }{2a_{NPi} (2n+1)\varepsilon _{0} } } $ is the dimensional factor, where $n$ is the general quantum and $m$ is the magnetic quantum numbers~\cite{8}, and $a_{NPi} $ is the radius of i-th NP.

Next, the near field of a single QD is written as
\begin{equation}
\hat{\textrm{E} }_{QDi} =\frac{1}{4\pi \varepsilon _{0} } \cdot \frac{3\vec{n}\left(\vec{n}\cdot \vec{e}_{QDi} \right)-\vec{e}_{QDi}}{r^{3}} \hat{d}_{QDi},
\label{eq:1}
\end{equation}
where the dipole moment operator is $\hat{d}_{QDi} =\mu _{QDi} \left(\hat{S}_{i} +\hat{S}_{i}^{+} \right)\vec{e}_{QD} $ expressed via creation operator $\hat{S}_{i}^{+} ={\left| e \right\rangle} _{i} {\left\langle g \right|} _{i} $ and annihilation operator $\hat{S}_{i}^{} ={\left| g \right\rangle} _{i} {\left\langle e \right|} _{i} $ of excitons and the dipole moment $\mu_{QDi} $ corresponding to interband transitions in QD, where ${\left| e \right\rangle} _{i} $ corresponds to exited and ${\left| g \right\rangle} _{i} $ ground states of the system. The presented operators satisfy commutation relationships $\left[\hat{S}_{i}^{+},\hat{S}_{i} \right]=D_{i} $ and $\left[\hat{S}_{i},D_{i} \right]=2\hat{S}_{i} $, where $D_{i} =\hat{S}_{i}^{+} \hat{S}_{i}-\hat{S}_{i} \hat{S}_{i}^{+} $ is the population imbalance operator; $\vec{e}_{QDi}$ determines the QD dipole moment orientation.

In conditions $\lambda_{1,2}\gg r>a_{NPi},a_{QDi} $ where $a_{QDi} $ is the QD radius, and $\lambda_{1,2} $ are the wavelengths of interband transition in the QDs, all pairwise interactions correspond to a dipole-dipole energy exchange. In particular, a Hamiltonian interaction between NP and QD can be written in the form $V_{i}^{QN} =-\hat{{\textrm{E}} }_{NPi}^{\parallel} \hat{d}_{QDi} $. The geometry presented in Fig.~\ref{fig:1} corresponds to the case $\theta =0$ and therefore the Legendre polynomials take the form $P_{1}^{0} \left(\cos \theta \right)=1$, $P_{1}^{m\ne 0} \left(\cos \theta \right)=0$. Thus, the near-field of $\textrm{NP}_{i}$ in the position of $\textrm{QD}_{i}$ location has the form
\begin{equation}
\hat{{\rm E} }_{NPi}^{\parallel} =\sqrt{\frac{\hbar \omega _{pi} a_{NPi}^{3} }{2\pi \varepsilon _{0} } } \frac{1}{r^{3} } \left(\hat{c}_{i} +\hat{c}_{i}^{+} \right)\vec{e}_{NPi},
\label{eq:2}
\end{equation}
where $\omega_{pi} $ determine plasmon frequencies of $\textrm{NP}_{i}$, $\varepsilon_{0}$ is vacuum permittivity. The interaction Hamiltonian for adjacent NPs $V^{NN} =-\hat{{\rm E} }_{NP1}^{\perp} \hat{d}_{NP2} $ is determined by orientation $\theta =\pi /2$, for which $P_{1}^{1} \left(\cos \theta \right)=1$ and $P_{1}^{m\ne 1} \left(\cos \theta \right)=0$. In this case, the expression for the field takes the form
\begin{equation}
\hat{{\rm E} }_{NPi}^{\perp} =\sqrt{\frac{\hbar \omega_{pi} a_{NPi}^{3} }{4\pi \varepsilon_{0} } } \frac{1}{r^{3} } \left(\hat{c}_{i} +\hat{c}_{i}^{+} \right)\vec{e}_{NPi}.
\label{eq:3}
\end{equation}
The dipole moment of $\textrm{NP}_{i}$ can be obtained from comparison (\ref{eq:3}) and the expression for the near field of $\textrm{NP}_{i}$ in analogy with (\ref{eq:1}). This expression has the form $\hat{d}_{NPi} =\mu_{NPi} \left(\hat{c}_{i} +\hat{c}_{i}^{+} \right)\vec{e}_{NP} $, where $\mu_{NPi} =\sqrt{4\pi \varepsilon _{0} \hbar \omega_{pi} a_{NPi}^{3} } $. Finally, the interaction Hamiltonian $V^{QQ} =-\hat{\textrm{E} }_{QD1} \hat{d}_{QD2} $ between two QDs in the spaser system is determined by the given geometry, for which $\left(\vec{n}\cdot \vec{e}_{QDi} \right)=0$.

Based on the necessity of internal symmetry in the layer arrangement for QD and NP in case of scaling the spaser system to a 2D array of spasers, we assume $r_{NP} =r_{QD}=r_{1}$, $r_{QN} =r$. The working regime of $2\times2$ spaser significantly depends on the ratio between frequencies $\omega_{1,2} $ of transitions in QDs and plasmon frequencies $\omega_{p1,p2} $. They are usually~\cite{6,28} considered almost equal to each other so that there are mostly linear plasmon interactions implemented in the system. Then, in the case $\omega_{i} \approx \omega_{pi} $, the corresponding Hamiltonian of interaction takes the form:
\begin{align}
\nonumber
H=&\hbar \omega_{p1} \hat{c}_{1}^{+} \hat{c}_{1} +\hbar \omega_{p2} \hat{c}_{2}^{+} \hat{c}_{2} +\frac{\hbar \omega_{1} }{2} D_{1} +\frac{\hbar \omega_{2} }{2} D_{2} \\
\nonumber
&+\hbar \Omega_{1} \left(\hat{c}_{1} \hat{S}_{1}^{+} + \hat{c}_{1}^{+} \hat{S}_{1}^{} \right) + \hbar \Omega_{2} \left(\hat{c}_{2} \hat{S}_{2}^{+} +\hat{c}_{2}^{+} \hat{S}_{2} \right) \\
\label{eq:4}
&+\hbar \Omega_{QQ} \left(\hat{S}_{1} \hat{S}_{2}^{+} +\hat{S}_{1}^{+} \hat{S}_{2} \right)+\hbar \Omega_{pp} \left(\hat{c}_{1} \hat{c}_{2}^{+} +\hat{c}_{1}^{+} \hat{c}_{2} \right),
\end{align}
where the 5th and 6th terms with $\Omega_{i} =\sqrt{\frac{\omega_{pi} a_{NPi}^{3} }{2\pi \varepsilon _{0} \hbar } } \frac{\mu_{QD} }{r^{3} } $ correspond to $V_{i}^{QN} $, the 7th with $\Omega_{QQ} =\frac{\mu_{QD}^{2} }{4\pi \varepsilon _{0} \hbar r_{1}^{3} } $ appears from $V^{QQ} $ and term with $\Omega_{pp} =\frac{\mu_{NP}^{2} }{4\pi \varepsilon _{0} \hbar r_{1}^{3} } $ is defined by Hamiltonian $V^{NN} $. We ignore cross-interaction between NP and QD placed diagonally. In (\ref{eq:4}), we will not take into account the Ferster transfer of energy between QDs, although in a real situation this must be done.

In this section, we will simulate the dynamics of our spaser system, using an average of parameters taken from a body of experimental work in this research area. As a model, we choose a spaser consisting of a gold NP and QD based on semiconductor CdSe~\cite{6ref,8ref}, for which the chosen states ${\left| g \right\rangle}_{i} $ and ${\left| e \right\rangle}_{i} $ correspond to hole level $1S\left(h\right)$ in the valence band and electron level $1S\left(e\right)$ in the conduction band. QD sizes can be estimated based on the plasmon mode frequency $\omega_{p} =\omega_{p1} =\omega_{p2} $, which for a spherical gold NP corresponds to a wavelength of $520 \; \textrm{nm}$~\cite{theor1}. To satisfy the exact resonance condition $\omega_{p} =\omega $ between the NP and QD, the size of the QD is approximately determined by the known dependence of the interband transition between the energy levels $1S\left(e\right)$ and $1S\left(h\right)$~\cite{grun} on its diameter:
\begin{align}
\label{eq:5}
E_{bb} = \hbar \omega = E_{g} +2\frac{\hbar^{2} \pi^{2} }{D_{QD}^{2} } \left(\frac{1}{m_{e} } +\frac{1}{m_{h} } \right),
\end{align}
where $D_{QD} =2a_{QD}$, $e$ is the electron charge, $\hbar $ is the Planck constant, $m_{e} $ and $m_{h} $ are the effective masses of the electron and hole in the bulk of the QD material with dielectric permittivity $\varepsilon $ and bandgap $E_{g} $ respectively. For CdSe, the corresponding parameters are $E_{g} /e=1.76 \; \textrm{eV}$, $m_{e} =0.125m_{0} $, $m_{h} =0.43m_{0} $ and $\varepsilon =10$. Using formula (\ref{eq:5}), we get $D_{QD} =4.97 \; \textrm{nm}$ for these parameters and $\lambda=520 \; \textrm{nm}$. The Bohr radius of exciton $R_{ex}$ for CdSe is $4.55 \; \textrm{nm}$~\cite{31}, therefore a strong confinement regime~\cite{32} will be observed for the QD excitons. The energy sublevels of the conductivity zone will be essentially separated for considered conditions. Therefore, the two-level model will be valid for QDs.

The dipole moment value of the interband transition~\cite{33} for CdSe is equal to $\mu_{QD} =0.309\cdot 10^{-28} \; \textrm{C}\cdot \textrm{m}$ in selected conditions. The dipole moment of the NP with a radius exactly matching the radius of the QD ($a_{NP}=a_{QD}$) is equal to $\mu_{NP} =4.548\cdot 10^{-28} \; \textrm{C}\cdot \textrm{m}$. We suggest that spaser $2\times 2$ is a square with characteristic sizes $r_{1} =r=5.3 \; \textrm{nm}$, the corresponding Rabi frequencies in (\ref{eq:4}) take the value $\Omega_{1} =\Omega_{2} =\Omega =2.026\cdot 10^{13} \; \textrm{s}^{-\textrm{1}} $, $\Omega_{QQ} =5.49\cdot 10^{11} \; \textrm{s}^{-\textrm{1}} $, $\Omega_{pp} =1.19\cdot 10^{14} \; \textrm{s}^{-\textrm{1}} $. We notice that the efficiency of dipole-dipole interaction between QDs in the presented geometry is significantly lower for similar efficiency both between the adjacent NPs and in an NP-QD pair. Thus, for further consideration, the term with $\Omega_{QQ} $ can be neglected, and we can proceed to the consideration of the following Heisenberg-Langevin system of equations obtained from (\ref{eq:4}):
\begin{subequations}
\label{eq:6}
\begin{eqnarray}
\dot{\hat{c}}_{1} &=& i\left(\Delta_{1} +\frac{i}{\tau_{c1} } \right)\hat{c}_{1} -i\Omega_{1} \hat{S}_{1} -i\Omega_{pp} \hat{c}_{2} +\hat{F}_{c1}, \\
\dot{\hat{c}}_{2} &=& i\left(\Delta_{2} +\frac{i}{\tau_{c2} } \right)\hat{c}_{2} -i\Omega_{2} \hat{S}_{2} -i\Omega_{pp} \hat{c}_{1} +\hat{F}_{c2}, \\
\dot{\hat{S}}_{1} &=& i\left(\delta_{1} +\frac{i}{\tau_{S1} } \right)\hat{S}_{1} +i\Omega_{1} D_{1} \hat{c}_{1} +\hat{F}_{S1}, \\
\dot{\hat{S}}_{2} &=& i\left(\delta_{2} +\frac{i}{\tau_{S2} } \right)\hat{S}_{2} +i\Omega_{2} D_{2} \hat{c}_{2} +\hat{F}_{S2}, \\
\dot{D}_{1} &=& -2i\Omega_{1} \left(\hat{S}_{1}^{+} \hat{c}_{1} -\hat{S}_{1} \hat{c}_{1}^{+} \right)-\frac{D_{1} -D_{01} }{\tau_{D1} } +\hat{F}_{D1}, \\
\dot{D}_{2} &=& -2i\Omega_{2} \left(\hat{S}_{2}^{+} \hat{c}_{2} -\hat{S}_{2} \hat{c}_{2}^{+} \right)-\frac{D_{2} -D_{02} }{\tau_{D2} } +\hat{F}_{D2},
\end{eqnarray}
\end{subequations}
where $\Delta_{1} =\bar{\omega }-\omega_{p1} $, $\Delta_{2} =\bar{\omega }-\omega_{p2} $, $\delta_{1} =\bar{\omega }-\omega_{1} $, $\delta_{2} =\bar{\omega }-\omega_{2} $, and parameters $\bar{\omega }$ and $D_{01(02)} $ correspond to the frequency and value of the spaser pumping respectively. In the deriving system (\ref{eq:6}) we used the rotating-wave approximation $\hat{c}=\hat{c}\cdot \exp \left(-i\bar{\omega }t\right)$ and $\hat{S}=\hat{S}\cdot \exp \left(-i\bar{\omega }t\right)$ upon passage to the new slow-varying operators $\hat{c}\left(\hat{c}^{+} \right)$ and $\hat{S}\left(\hat{S}^{+} \right)$. In equations (\ref{eq:6}) the characteristic parameters of the decay rate for plasmons $\frac{1}{\tau_{c1(c2)} } $ in NP, the decay rate of excitons $\frac{1}{\tau_{S1 \left(S2 \right)} } $ in excited QDs, and also the operators of Langevin noises $\hat{F}_{c1(c2)} $ ($\hat{F}_{S1(S2 )} $, $\hat{F}_{D1(D2)} $) are introduced phenomenologically~\cite{34}, proceeding from the conditions of system interaction with the reservoir. The pump operator and appropriate time are $D_{01(02)} =\frac{2\tau_{S1(S2 )} -\tau_{p1(p2)} }{2\tau_{S1(S2 )} +\tau_{p1(p2)} } \hat{I}$ and $\tau_{D1(D2)} =\left(\frac{1}{2\tau_{S1(S2 )} } +\frac{1}{\tau_{p1(p2 )} } \right)^{-1} $~\cite{35} respectively. Here the parameters $\tau_{p1(p2)} $ are characteristic pumping times. Furthermore, we assume that both NPs (and also both QDs) are identical to each other, i.e. $\Omega_{1} =\Omega_{2} $, $\tau_{c1} =\tau_{c2} =\tau_{c} $, $\tau_{S1} =\tau_{S2} =\tau_{S} $, $\Delta_{1} =\Delta_{2} =\Delta $ ($\omega_{p1} =\omega_{p2} =\omega_{p} $), $\delta_{1} =\delta_{2} =\delta $ ($\omega_{1} =\omega_{2} =\omega $). In addition, we assume that the populations in both QDs change synchronously, i.e. $D_{1} =D_{2} =D$.

After the material, the size of the nano-objects and their position in the spaser structure have been defined, we try to fulfill preliminary optimization of the system by means of varying the dissipative system parameters. We use these parameters because the decay rates of excitations have a significant dependence on the chemical composition~\cite{37} and the purity of the nano-objects' surface~\cite{38} and can be varied within a wide range. Furthermore, the experimenter always has the ability to control the system's pumping rate. The purpose of our preliminary optimization is to obtain time-independent solutions for the average number of plasmons and excitons generated in the spaser system.

Moving on to the estimates of relaxation parameters of the problem, it should be noted that the decay rate of plasmon mode $\gamma_{p} =\frac{1}{\tau_{c} } =\frac{1}{\tau_{J} } +\frac{1}{\tau_{R} } $ is determined by the characteristic time of radiation $\tau_{R} $ and Joule $\tau_{J} $ losses. However, under the condition $\frac{1}{\tau_{J} } \approx 30\frac{1}{\tau_{R} } $~\cite{36} we do not have to take radiation losses into account and Joule losses are normally determined by electron collisional frequency in metal $\gamma_{s}$, i.e. $\gamma_{p} \approx \gamma_{s} $. For gold, we can take the value $\gamma_{s} =4\cdot 10^{13} \; \textrm{s}^{-\textrm{1}} $.

The parameter $\frac{1}{\tau_{S} } =\frac{1}{\tau_{R} } +\frac{1}{\tau_{F} } $ represents the total rate of radiative (with time $\tau_{R} $) and nonradiative (with time $\tau_{F} $) losses in the QD. At the same time, the parameter $\frac{1}{\tau_{F} } $ gives the main contribution, since the processes of nonradiative recombination of excitons (with exciting phonon modes) occur in short times. Note that, annealing technology~\cite{38,39} and the use of core-shell QDs~\cite{6ref,8ref} allow a significant increase in $\tau_{F}$. Additional influence on the rate of decay is provided by the metallic NP approaching the QD. Following the article~\cite{6}, we accept $\frac{1}{\tau_{S} } =4\cdot 10^{10} \; \textrm{s}^{-\textrm{1}} $. The characteristic time of the dipole-dipole interactions between NP and QD is in the order of $0.5\cdot10^{-4} \; \textrm{ns}$, which is much less than the characteristic rate of exciton decay in the QD and is comparable to decay rate of plasmons. This situation corresponds to a strong coupling regime (see~\cite{3ref}).

Then we determine the characteristic spaser generation frequency (spasing frequency) $\bar{\omega }$, spasing threshold $D_{th} $, and also find the possible stationary regimes of its time evolution. To achieve this purpose, at the initial stage we write a system of equations for average values, analogical to an operator system (\ref{eq:6}), assuming that $c=\left\langle \hat{c}\right\rangle $, $c^{*} =\left\langle \hat{c}^{+} \right\rangle $ and Langevin noise operators are equal to zero. Then, assuming $\dot{c}_{1(2)} =\dot{D}=\dot{S}_{1(2)} =0$, we express $c_{2} $ from an equation (\ref{eq:6}a) and substitute the obtained expression in the equations of average values (\ref{eq:6}b) and (\ref{eq:6}d). After this, we express the parameter $S_{2} $ from (\ref{eq:6}d) and substitute it in (\ref{eq:6}b) and (\ref{eq:6}c). The resulting new system of algebraic equations on the average values will be written in the form:
\begin{subequations}
\label{eq:7}
\begin{eqnarray}
\Omega Dc_{1} +BS_{1} &=&0, \\
\nonumber
\left(\frac{A^{2} }{\Omega _{pp} } +\frac{A\Omega ^{2} D}{B\Omega _{pp} } -\Omega _{pp} \right)c_{1} \mspace{28mu}&& \\
+ \left(-\frac{\Omega A}{\Omega _{pp} } -\frac{\Omega ^{3} D}{\Omega _{pp} B} \right)S_{1} &=&0,
\end{eqnarray}
\end{subequations}
where the new definitions $A=A_{R} +iA_{I} =\Delta +\frac{i}{\tau_{c} } $ and $B=B_{R} +iB_{I} =\delta +\frac{i}{\tau_{S} } $ are introduced. The system (\ref{eq:7}) can have non-trivial solutions in the case when the matrix determinant for the left side of the system of equations is equal to zero. Thus, we obtain a system of two self-consistent equations for the real and imaginary parts of said determinant
\begin{subequations}
\label{eq:8}
\begin{eqnarray}
\nonumber
\Omega^{4} D^{2} +2\left(A_{R} B_{R} -A_{I} B_{I} \right)\Omega^{2} D \mspace{43mu}&&\\
\nonumber
+ \left(B_{I}^{2} -B_{R}^{2} \right)\left(A_{I}^{2} -A_{R}^{2} +\Omega_{pp}^{2} \right) && \\
- 4A_{I} A_{R} B_{I} B_{R} &=&0, \\
\nonumber
2\left(A_{R} B_{I} +A_{I} B_{R} \right)\left(A_{R} B_{R} -A_{I} B_{I} +\Omega^{2} D\right) && \\
- 2\Omega_{pp}^{2} B_{I} B_{R} &=&0.
\end{eqnarray}
\end{subequations}

The solutions of system (\ref{eq:8}) determine the spasing frequency $\bar{\omega}$ and the threshold $D_{th}$. This system of equations has two roots
\begin{subequations}
\label{eq:9}
\begin{eqnarray}
\bar{\omega }_{\pm } &=&\frac{\tau_{S} \omega +\tau_{c} \left(\omega_{p} \pm \Omega_{pp} \right)}{\tau_{c} +\tau_{S} }, \\
D_{th,\mp } &=&\frac{1+\left(\frac{\tau_{c} \tau_{S} }{\tau_{c} +\tau_{S} } \right)^{2} \left(\omega -\omega_{p} \mp \Omega_{pp} \right)^{2} }{\tau_{c} \tau_{S} \Omega^{2} }
\end{eqnarray}
\end{subequations}
one of which ($\bar{\omega}_{-} ,D_{th,+} $) will give an unstable solution for (\ref{eq:6}). Therefore, we will assume that $D_{th} =D_{th,-} $.

We note that in the case $\Omega_{pp} =0$ solutions (\ref{eq:9}) correspond to the known model of spaser $1\times 1$ which consists of one QD and one NP~\cite{7}. However, the presence of a near-field interaction between two NPs significantly increases the spasing threshold. Note that the solutions obtained in (\ref{eq:9}) correspond only to linear energy exchange between the NP and QD determined by Hamiltonian (\ref{eq:4}). Under the selected parameters of interaction, the threshold values are $D_{\textrm{th}}^{\textrm{1}\times \textrm{1}} =D_{th} \left(\Omega_{\textrm{pp}} =0\right)=0.0039$ and $D_{\textrm{th}}^{\textrm{2}\times \textrm{2}} =D_{th} \left(\Omega_{\textrm{pp}} =1.19\cdot 10^{14} \; \textrm{s}^{-\textrm{1}} \right)=0.0383$. In further simulation, the pumps are chosen according to the conditions $D_{01}=D_{02}=D_{0} > \max (D_{\textrm{th}}^{\textrm{1}\times \textrm{1}}, D_{\textrm{th}}^{\textrm{2}\times \textrm{2}} )$. Finally, we choose the value $D_{0} = 0.1$.
\begin{figure}[t]
\includegraphics[width=0.8\columnwidth]{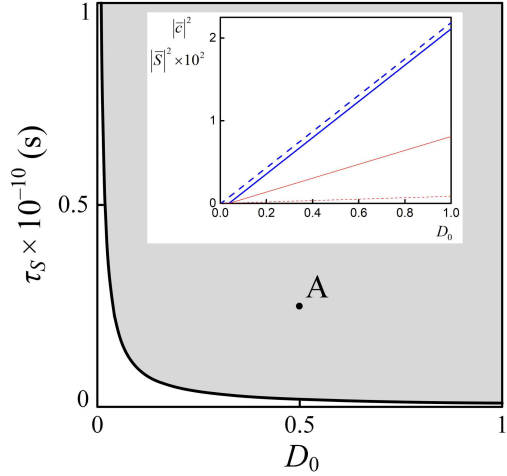}
\caption{\label{fig:2} (Color online) The parametric plane (pump value $D_{0} $, decay time of excitons in QD $\tau_{S} $) with the depicted stability area of the $2\times 2$ spaser and a point A ($0.5$, $0.25\cdot 10^{-10} \; \textrm{s}$). In the inset: the dependence of average numbers of plasmons $\left|\bar{c}\right|^{2} $ (thick blue lines) and excitons $\left|\bar{S}\right|^{2} $ (thin red lines) on the value of the pumping for $2\times 2$ spaser with accounting for $\Omega_{pp} $ (solid lines) and for the $1\times 1$ spaser without accounting for $\Omega_{pp} $ (dashed lines). Parameters of interaction: $\omega_{p} =\omega =3.625\cdot 10^{15} \; \textrm{s}^{-\textrm{1}} $, $\Omega =2.026\cdot 10^{13} \; \textrm{s}^{-\textrm{1}} $, $\Omega_{pp} =1.19\cdot 10^{14} \; \textrm{s}^{-\textrm{1}} $, $\tau_{c} =0.25\cdot 10^{-13} \; \textrm{s}$, $\tau_{D} =2.85\cdot 10^{-15} \; \textrm{s}$.}
\end{figure}
\begin{figure}[t]
\includegraphics[width=\columnwidth]{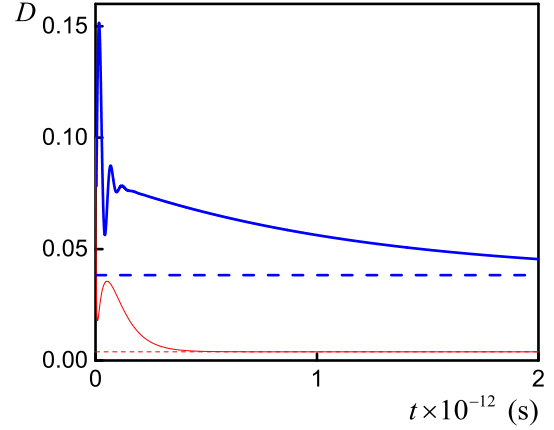}
\caption{\label{fig:3} (Color online) The time dependence of population imbalance parameter $D$ for $1\times 1$ spasers (thin red lines) and $2\times 2$ (thick blue lines), calculated by using formulas (\ref{eq:9}) (dashed lines) and by using direct numerical simulation (solid lines) of the system (\ref{eq:6}). Initial values: $c\left(0\right)=\bar{c} \cdot 1.05$, $S\left(0\right)=\bar{S} \cdot 0.95$, $D\left(0\right)=0.1$, where steady-state solutions $\bar{c} =0.7375+0.7375i$, $\bar{S} =-1.4562+1.4562i$ for $1\times 1$ spaser and $\bar{c} =0.7115+0.7115i$, $\bar{S} =-5.5792-2.7696i$ for $2\times 2$ spaser with $\phi \left(0\right)=\pi /4$. The simulation parameters correspond to point A from Fig.~\ref{fig:2}.}
\end{figure}

Assuming $c_{1} =c_{2} =c$ ($S_{1} =S_{2} =S$), we express $c$ from equation (\ref{eq:7}b) and substitute it in the result of summing the equations (\ref{eq:6}e) and (\ref{eq:6}f) under condition $\dot{D}=0$. As a result, steady-state solutions for the amplitudes of plasmons and excitons take forms which are determined up to the phase $\phi $:
\begin{subequations}
\label{eq:10}
\begin{eqnarray}
\bar{c} &=&e^{i\phi_{i} } \sqrt{\frac{\tau_{c} }{4\tau_{D} } \left(D_{0} -D_{th} \right)}, \\
\bar{S} &=&\frac{i+\frac{\tau_{c} \tau_{S} }{\tau_{c} +\tau_{S} } \left(\omega -\omega_{p} -\Omega_{pp} \right)}{\tau_{c} \Omega } \bar{c}.
\end{eqnarray}
\end{subequations}

A test on the stability of the obtained solutions was carried out by analysis of eigenvalues $\lambda_{i} $ of the linearized system of equations on average values (\ref{eq:6}) near fixed points (\ref{eq:10}), as well as by using a direct numerical simulation of this system. We found that for the obtained solutions (\ref{eq:10}), one of the roots of the characteristic equation for the linearized system (\ref{eq:6}) on average values is always equal to zero, while other eigenvalues are satisfied to inequality $\textrm{Re}\left(\lambda_{i} \right)<0$. Thus, the system moves to the boundary of aperiodic stability in the absence of external synchronization. From a mathematical point of view, in this case further analysis of the full nonlinear system (\ref{eq:6}) should be carried out. Nevertheless, numerical analysis of the system (\ref{eq:6}) demonstrates the stability of obtained solutions (\ref{eq:10}).

The parametric plane formed by the combination of the pump value $D_{0}$ and the characteristic time of excitons decay $\tau_{S} $ with the depicted stability area for the solutions (\ref{eq:10}) verified by numerical simulation of the system (\ref{eq:6}) is presented in Fig.~\ref{fig:2}. This area results from optimizing the parameters of our system. Obviously, the impurities in material of QDs lead to the decreasing of $\tau_{S}$. Therefore, it is necessary to increase the pump value $D_{0} $ to maintain stationary conditions for spasing. At the same time, increasing $D_{0}$ leads to the linear growth of the number of plasmons $\left|\bar{c}\right|^{2}$, inset in Fig.~\ref{fig:2}.

\begin{figure}[t]
\includegraphics[width=\columnwidth]{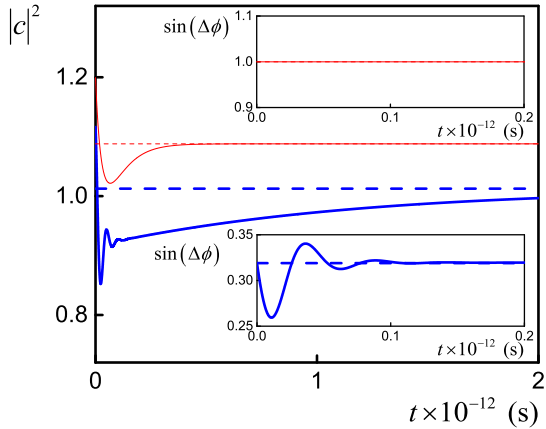}
\caption{\label{fig:4} (Color online) The time dependence of the average number of plasmons $\left|c\right|^{2} $ for the $1\times 1$ spaser (thin red lines) and for the $2\times 2$ spaser (thick blue lines), calculated using formulas (\ref{eq:10}a) (dashed lines) and by using direct numerical simulation (solid line) of the system (\ref{eq:6}). Insets: dependence of the relative phase between plasmon and exciton modes for the $2\times 2 $ spaser (bottom inset) and for the $1\times 1 $ spaser (top inset). The initial values and parameters of the simulation correspond to Fig.~\ref{eq:3}.}
\end{figure}

In Fig.~\ref{fig:3}, performed for the parameter $D$, the agreement between analytical and numerical results is clearly observed when choosing the parameters of interaction for point A from the presented area of stability in Fig.~\ref{fig:2}. At the same time, the dynamics of the transition process to the stationary values for $D$ and $\left|c\right|^{2} $ has a significant dependence on $\Omega_{pp} $. In particular, the rapid synchronization of energy exchange processes between QD and NP under the condition $\Omega_{pp} =0$ ($1\times1$ spaser) is replaced by a longer stabilization process, taking into account the contribution of $\Omega_{pp} $ ($2\times2$ spaser), in Figs.~\ref{fig:3},~\ref{fig:4}. It is due to a more intensive exchange of energies in the spaser $2\times2$. This can be observed by using the temporal dynamics of $\sin \left(\Delta \phi \left(t\right)\right)$, where the parameter $\Delta \phi \left(t\right)=\textrm{Arg}\left(\frac{S\left(t\right)}{c\left(t\right)} \right)$ determines a relative phase between plasmons and excitons (see insets in Fig.~\ref{fig:4}). Taking into account $\Omega_{pp}$, the steady-state solution of $\Delta \phi_{\textrm{st}}$ significantly differs from $\Delta \phi_{\textrm{st}} =\pi /2$ for the case $\Omega_{pp} =0$~\cite{7}. At the same time, the amount $\Delta \phi_{\textrm{st}} $ is independent of the initial values of $c\left(0\right)$ and $S\left(0\right)$.

We will now proceed to consider the spasers quantum statistical properties~\cite{40,40a} on the basis of the study the second-order autocorrelation function $G^{\left(2\right)}$ for plasmons in the form
\begin{equation}
\label{eq:11}
G_{i}^{\left(2\right)} \left(t,\tau\right) = \frac{\left\langle \hat{c}_{i}^{+} \left(t\right) \hat{c}_{i} \left(t\right) \hat{c}_{i}^{+} \left(t+\tau\right) \hat{c}_{i} \left(t+\tau\right) \right\rangle }{\left\langle \hat{c}_{i}^{+} \left(t\right) \hat{c}_{i} \left(t\right)\right\rangle \left\langle \hat{c}_{i}^{+} \left(t+\tau\right) \hat{c}_{i} \left(t+\tau\right)\right\rangle}-1.
\end{equation}

In our case the parameter $G_{i}^{\left(2\right)} \equiv G_{i}^{\left(2\right)}\left(t,0\right)$ indicates a bunching effect if $G_{i}^{\left(2\right)}>0$ (super-Poisson statistics), or an antibunching effect if $G_{i}^{\left(2\right)} <0$ (nonclassical states with sub-Poisson statistics) of plasmons.
\begin{figure}[t]
\includegraphics[width=\columnwidth]{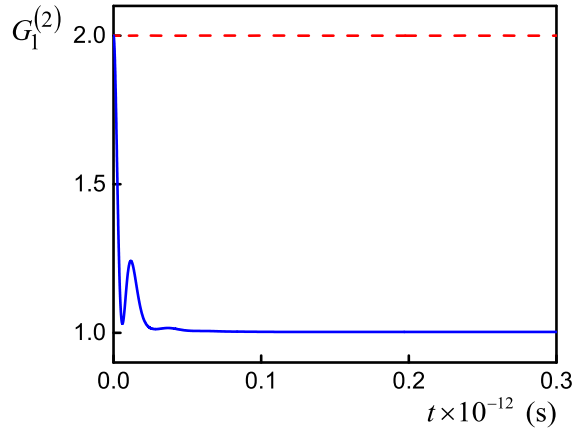}
\caption{\label{fig:5} (Color online) The time dependence of the autocorrelation function $G_{1}^{\left(2\right)} $ for the $2\times 2$ spaser (solid blue line) and for the $1\times 1$ spaser (dashed red line) with parameters of interaction corresponding to Fig.~\ref{fig:3}.}
\end{figure}

Figure~\ref{fig:5} presents the results of numerical solution of the system (\ref{eq:A}) for calculation of $G_{1}^{\left(2\right)} \left(t\right)$ for one plasmon mode in the Fock state and choosing the same parameters of interaction as for point A from Fig.~\ref{fig:2}.

Initial second-order correlators for plasmons and for excitons taken within the same mode are replaced by the product of their average values, i.e.
\begin{eqnarray}
\nonumber
&\left.\left\langle\hat{S}_{i}^{+}\hat{S}_{i}\right\rangle\right|_{t=0}=\left|S_{i}\left(0\right)\right|^{2},& \\
\nonumber
&\left.\left\langle\hat{S}_{i}\hat{S}_{i}^{+}\right\rangle\right|_{t=0}=\left|S_{i}\left(0\right)\right|^{2}-D_{i}\left(0\right),& \\
\nonumber
&\left.\left\langle \hat{S}_{i}^{2} \right\rangle\right|_{t=0}=S_{i}^{2}\left(0\right), \; \left.\left\langle \left(\hat{S}_{i}^{+}\right)^{2} \right\rangle\right|_{t=0}=\left(S_{i}^{*}\left(0\right)\right)^{2};& \\
\nonumber
&\left.\left\langle\hat{c}_{i}^{+}\hat{c}_{i}\right\rangle\right|_{t=0}=\left|c_{i}\left(0\right)\right|^{2}, \; \left.\left\langle\hat{c}_{i}\hat{c}_{i}^{+}\right\rangle\right|_{t=0}=1+\left|c_{i}\left(0\right)\right|^{2},& \\
\nonumber
&\left.\left\langle \hat{c}_{i}^{2} \right\rangle\right|_{t=0}=c_{i}^{2}\left(0\right), \; \left.\left\langle \left(\hat{c}_{i}^{+}\right)^{2} \right\rangle\right|_{t=0}=\left(c_{i}^{*}\left(0\right)\right)^{2},&
\end{eqnarray}
where $i\in\left\{1,2\right\}$. However, initial values for second-order correlators of plasmons and excitons taken from different modes are assumed to equal zero, and the initial population imbalance is $D_{1,2} \left(0\right)=0.1$. We expanded a four-order correlator in (\ref{eq:11}) by using Wick's theorem and found that $G_{1}^{\left(2\right)} \left(0\right)=2$. This situation corresponds to the initial super-Poisson statistics of plasmons.

During the process of stabilizing the $1\times 1$ spaser (dashed line in Fig.~\ref{fig:5}), changes in quantum statistics are minimal. In contrast to this, intensive energy exchange between a pair of NPs in the $2\times2$ spaser leads to a significant decrease in the value of $G_{1}^{\left(2\right)} \left(t\right)$ (see solid line in Fig.~\ref{fig:5}). However, even if we choose any arbitrary values for the control parameters, the values of the autocorrelation function $G_{1}^{\left(2\right)} \left(t\right)$ never fall below zero, which characterizes coherent plasmon source. Thus, we conclude that the presence of a strong dipole-dipole interactions between NPs significantly changes the initial super-Poisson statistic of localized plasmons in the spaser system. In addition, the analysis of the cross-correlation function $G_{12}^{\left(2\right)} \left(t\right)$ for plasmons demonstrates its purely classical behavior due to the absence of nonlinear interaction in the system.
\begin{figure*}[t]
\includegraphics[width=1.5\columnwidth]{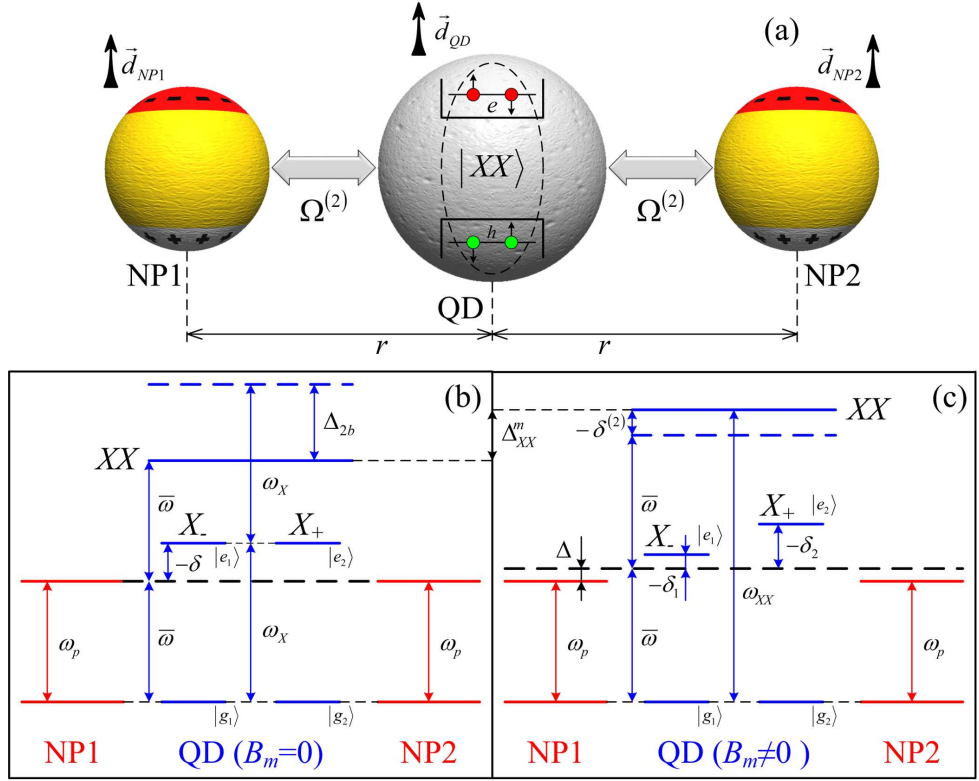}
\caption{\label{fig:6} (a) A model of a three-particle spaser consisting of two NPs and single QD (NP-QD-NP), in which the generation of entangled plasmons is realized due to the QD's biexciton states decay. The mapping of exciton and plasmon energy levels is in the absence of (b) and in the presence of (c) external magnetic field ($\Delta_{XX}^{m}=2kB_{m}^{2}$).}
\end{figure*}

\section{\label{sec3}Entangled plasmon generation in the nonlinear regime of a three-particle spaser controlled by external magnetic field}
In this section we consider nonlinear plasmon-exciton interactions, and also perform a complete optimization of the spaser system parameters in order to demonstrate the formation of nonclassical states of localized plasmons.
Nonlinear regimes of interaction between NP and QD can be realized, firstly, in the presence of a two-photon pump in the system~\cite{41,22} and secondly, under the condition that the coupling energy between two electron-hole pairs is of the same order of magnitude as the internal coupling energy of a single pair. In this situation the coupled states of two electron-hole pairs (biexcitons of QDs) can appear~\cite{41}. The energy of biexciton state $XX$ differs from the double energy of the exciton $X$ by the biexciton binding energy $\Delta_{2b}$ (see Fig.~\ref{fig:6}).

The decay of the biexciton state $XX$ of the QD can occur in various scenarios. The first variant is cascade processes~\cite{ihara}, when the biexciton $XX$ splits into the $X$ exciton state with the emission of a single photon, and then the recombination in the $X$ exciton occurs with the appearance of a second photon~\cite{S1}.

However, for a QD without spatial symmetry, the energy of the intermediate $X$ state will depend on the spin state of the electron in the conduction band. This is a well-known effect of fine structure splitting (FSS). If the energy $E_{FSS}$ is greater than the width of the emission line, then the energies of the pair of photons that are generated during the decay through level $X_{+}$ significantly differ from the energy of the photons produced during the decay of the biexciton through the intermediate state $X_{-}$. Such a case is not interesting for quantum informatics.

If $E_{FSS}$ is much smaller than the line width, then the photons generated in each of the cascades ($XX \rightarrow X_{+} \rightarrow g$ and $XX \rightarrow X_{-} \rightarrow g$) are indistinguishable in energy (see Fig.~\ref{fig:6}) and there is a so-called witch patch interference that results in entanglement between photons~\cite{S1}. However, it is very difficult to control the quantum statistics and the correlation properties of the photons directly during the experiment with this approach.

Therefore, we choose another regime in which a cascade process is not realized in the system of QDs, but a pure two-quantum transition occurs~\cite{S2}. In the conditions corresponding to scheme in Fig.~\ref{fig:6} this regime is realized due to far-off-resonant interaction in the case $\left|\delta\right|>\Omega_{1,2}$. Such a process can lead to the generation of nonclassical states of plasmons and it is described by the following Hamiltonian~\cite{S3}:
\begin{eqnarray}
\nonumber
H&=&\hbar \omega_{p1} \hat{c}_{1}^{+} \hat{c}_{1} +\hbar \omega_{p2} \hat{c}_{2}^{+} \hat{c}_{2} +\frac{\hbar \omega_{XX} }{2} D \\
\label{eq:12}
&& +\hbar \Omega^{\left(2\right)} \Bigl(\hat{c}_{1} \hat{c}_{2} \hat{S}^{+} +\hat{c}_{1}^{+} \hat{c}_{2}^{+} \hat{S} \Bigr),
\end{eqnarray}
where the last term in the brackets comprises the annihilation operator $\hat{S}$ of the biexciton $XX$ state and the creation operators $\hat{c}_{1}^{+}$ and $\hat{c}_{2}^{+}$ of a pair of plasmons, whose energies differ slightly for the different intermediate levels $X_{+}$ ($X_{-}$) with frequencies $\omega_{X_{+}}$ ($\omega_{X_{-}}$). The basis states $\left|g\right\rangle_{1}=\left|1S\left(h\right), \; m_{s}=-1/2\right\rangle$, $\left|g\right\rangle_{2}=\left|1S\left(h\right), \; m_{s}=+1/2\right\rangle$, $\left|e\right\rangle_{1}=\left|1S\left(e\right), \; m_{s}=+1/2\right\rangle$, $\left|e\right\rangle_{2}=\left|1S\left(e\right), \; m_{s}=-1/2\right\rangle$ for electrons and holes of QDs in Fig.~\ref{fig:6} differ in the $m_{s}$ values. The parameter $\Omega^{\left(2\right)}=\frac{\Omega_{1}\Omega_{2}}{2\left|\delta\right|}$ is the effective two-quantum Rabi frequency. Parameters $\delta=\bar{\omega}-\omega_{X}$ and $\delta^{\left(2\right)}=2\bar{\omega}-\omega_{XX}$ are the effective detunings, where $\omega_{XX}=\omega_{X_{-}}+\omega_{X_{+}}-\Delta_{2b}$ is the biexciton frequency~\cite{S4}. We neglect the FSS and in the absence of an external magnetic field we assume $\omega_{X_{-}}=\omega_{X_{+}}=\omega_{X}$. We suppose that both NPs are identical, i.e. $\omega_{p1} =\omega_{p2} =\omega_{p} $, $\Delta=\bar{\omega}-\omega_{p}$, and the interaction between distant NPs placed at distance $2r$ can be ignored in the problem. The orientation of the dipole moments of the nano-objects in Fig.~\ref{fig:6} corresponds to the nonradiative regime of the spaser, therefore the Rabi frequency of interaction between the QD and NP is $\Omega_{1}=\Omega_{2}=\frac{\mu_{QD}}{r^{3}}\sqrt{\frac{\omega_{p}a^{3}_{NP}}{4 \pi \hbar \varepsilon_{0}}}$.

We use an approximation of slowly varying amplitudes, i.e. we assume that $\hat{c}_{1}=\hat{c}_{1}e^{-i\bar{\omega}t}$, $\hat{c}_{2}=\hat{c}_{2}e^{-i\bar{\omega}t}$ and $\hat{S}=\hat{S}e^{-2i\bar{\omega}t}$. Then the system of Heisenberg-Langevin equations corresponding to (\ref{eq:12}) can be represented in the following form:
\begin{subequations}
\label{eq:13}
\begin{eqnarray}
\dot{\hat{c}}_{1} &=&i\left(\Delta +\frac{i}{\tau_{c} } \right)\hat{c}_{1} -i\Omega^{\left(2\right)} \hat{c}_{2}^{+} \hat{S} + \hat{F}_{c1}, \\
\dot{\hat{c}}_{2} &=&i\left(\Delta +\frac{i}{\tau_{c} } \right)\hat{c}_{2} -i\Omega^{\left(2\right)} \hat{c}_{1}^{+} \hat{S} + \hat{F}_{c2}, \\
\dot{\hat{S}} &=&i\left(\delta^{\left(2\right)} +\frac{i}{\tau_{S} } \right)\hat{S} +i\Omega^{\left(2\right)} \hat{c}_{1} \hat{c}_{2} D + \hat{F}_{S}, \\
\nonumber
\dot{D} &=&2i\Omega^{\left(2\right)} \left(\hat{c}_{1}^{+} \hat{c}_{2}^{+} \hat{S} -\hat{S}^{+} \hat{c}_{1} \hat{c}_{2} \right) \\
&& \mspace{117mu}
- \frac{D -D_{0} }{\tau_{D} } +\hat{F}_{D},
\end{eqnarray}
\end{subequations}
where the decay rates of different plasmon modes are assumed to be equal each other, i.e. $\tau_{c1}=\tau_{c2}\equiv\tau_{c}$. We also assume that the pump $D_{0} $ has a rate of $1/\tau_{D} $ for the exciton mode.

Now we define the generation conditions for a nonlinear spaser. For this we replace the operators in the system (\ref{eq:13}) with C-numbers and obtain the system of algebraic equations for the stationary regime:
\begin{subequations}
\label{eq:14}
\begin{eqnarray}
0 &=&A_{1} c_{1} -i\Omega^{\left(2\right)} c_{2}^{*} S, \\
0 &=&A_{1} c_{2} -i\Omega^{\left(2\right)} c_{1}^{*} S, \\
0 &=&B_{1} S +i\Omega^{\left(2\right)} c_{1} c_{2} D, \\
0 &=&2i\Omega^{\left(2\right)} \left(c_{1}^{*} c_{2}^{*} S -S^{*} c_{1} c_{2} \right) - \frac{D -D_{0} }{\tau_{D} },
\end{eqnarray}
\end{subequations}
where we introduce the new parameters $A_{1} =i\Delta -\frac{1}{\tau_{c} } $ and $B_{1} =i\delta^{\left(2\right)} -\frac{1}{\tau_{S} } $. Equations for plasmon-exciton modes can be obtained by expressing $S $ from (\ref{eq:14}c) and substituting it in (\ref{eq:14}a) and (\ref{eq:14}b). Finally, these equations have the forms:
\begin{subequations}
\label{eq:15}
\begin{eqnarray}
A_{1} B_{1} -\left.\Omega^{\left(2\right)}\right.^{2} \left|c_{2}\right|^{2} D &=&0, \\
A_{1} B_{1} -\left.\Omega^{\left(2\right)}\right.^{2} \left|c_{1}\right|^{2} D &=&0.
\end{eqnarray}
\end{subequations}

The spasing frequency can be found from the equations (\ref{eq:15}a) and (\ref{eq:15}b), according to which the expression $A_{1} B_{1} $ must always take real values. Since $A_{1} B_{1} =\frac{1}{\tau_{c} \tau_{S} } -\Delta \delta^{\left(2\right)} -i\left(\frac{\Delta }{\tau_{S} } +\frac{\delta^{\left(2\right)} }{\tau_{c} } \right)$, that restriction is satisfied in the case
\begin{equation}
\label{eq:16}
\frac{\Delta }{\tau_{S} } =-\frac{\delta^{\left(2\right)} }{\tau_{c} },
\end{equation}
and along with that, $\textrm{Re}\left(A_{1} B_{1} \right)>0$. Relationship (\ref{eq:16}), in particular, will be satisfied under conditions $\delta^{\left(2\right)} <0$ and $\Delta >0$.

To determine the steady-state values $\bar{c}_{i} $, $\bar{S} $ and $\bar{D}$ it is necessary to perform calculations by analogy with Sec.~\ref{sec2}, but they will be applied to the newly defined coupled plasmon-exciton modes. In particular, from the equations (\ref{eq:14}a) and (\ref{eq:14}b) it is possible to derive the following expressions
\begin{subequations}
\label{eq:18}
\begin{eqnarray}
\bar{c}_{2}^{*}S &=&\frac{A_{1} \bar{c}_{1}}{i \Omega^{\left(2\right)}},\\
\bar{c}_{1}^{*}S &=&\frac{A_{1} \bar{c}_{2}}{i \Omega^{\left(2\right)}}.
\end{eqnarray}
\end{subequations}
After the substitution of these formulas in equation (\ref{eq:14}d) the amplitude of plasmons for a stationary solution takes a form similar to (\ref{eq:10}a):
\begin{equation}
\label{eq:19}
\bar{c}_{i} =\frac{e^{i\phi_{i} } }{2} \sqrt{\frac{\tau_{c} }{\tau_{D} } \left(D_{0} -\bar{D} \right)}.
\end{equation}
These stationary solutions are defined up to the plasmon phase $\phi_{i} $. After substitution of the formulas (\ref{eq:19}) in equation (\ref{eq:14}c) and expression of $\bar{S}$ from (\ref{eq:14}c), the amplitude of exciton mode takes the form
\begin{equation}
\label{eq:20}
\bar{S} =-\frac{i\Omega^{\left(2\right)} \tau_{c}}{4B_{1}\tau_{D}}\bar{D}\left(D_{0}-\bar{D}\right)e^{i\left(\phi_{1}+\phi_{2}\right)}.
\end{equation}

Having substituted (\ref{eq:19}) in (\ref{eq:15}) we obtain an equation for population imbalance in the form:
\begin{equation}
\label{eq:22}
\bar{D}^{2}-\bar{D}D_{0}+\frac{4\tau_{D}}{\left.\Omega^{\left(2\right)}\right.^{2}\tau_{c}}\left(\frac{1}{\tau_{c}\tau_{S}}-\Delta\delta^{\left(2\right)}\right)=0.
\end{equation}

The equation (\ref{eq:22}) has two roots, only one of which is stable. The corresponding stable solution takes the form:
\begin{equation}
\label{eq:24}
\bar{D} =\frac{D_{0}}{2} -\frac{1}{2} \sqrt{D_{0}^{2}+\frac{16\tau_{D}}{\left.\Omega^{\left(2\right)}\right.^{2}\tau_{c}}\left(\Delta\delta^{\left(2\right)}-\frac{1}{\tau_{c}\tau_{S}}\right)}.
\end{equation}

Initializing our system in the absence of a magnetic field, we assume $\bar{\omega}_{0}=\omega_{XX0}/2$, where $\omega_{XX0}=\left.\omega_{XX}\right|_{B_{m}=0}$. Then we get $\delta_{0}^{\left(2\right)}=0$ and the required significant value of detuning is $\delta_{0}= -\Delta_{2b}/2$. From (\ref{eq:16}) we also obtain an additional condition $\Delta=0$. Then the frequency of the transition in the QD can be expressed as $\omega_{X0}=\omega_{p}+\Delta_{2b}/2$. We again choose $\lambda_{p}=520 \; \textrm{nm}$ for gold and $\Delta_{2b}=2.881 \cdot 10^{14} \; \textrm{s}^{-\textrm{1}} \; \left(0.19 \; \textrm{eV}\right)$ for the CdSe QD~\cite{binding}. Using the expression (\ref{eq:5}), we determine the size of the QD $D_{QD}=4.635 \; \textrm{nm}$ to satisfy the conditions of two-quantum transition in the scheme. The dipole moment of corresponding transition in the QD will take new value $\mu_{QD} =0.303\cdot 10^{-28} \; \textrm{C}\cdot \textrm{m}$.

Now we need to optimize the geometry of the spaser system and the dissipative parameters of the nano-objects in order to obtain stable stationary solutions. The simulation parameters correspond to $\tau_{c} =5 \cdot 10^{-12} \; \textrm{s}$, $\tau_{S} =4 \cdot 10^{-11} \; \textrm{s}$ and frequency detuning to $\delta_{0}=-1.441 \cdot 10^{14} \; \textrm{s}^{-\textrm{1}}$. The distance between the NP and QD equals $r=5 \; \textrm{nm}$. For a NP radius $a_{NP}=D_{QD}/2$, the single-plasmon Rabi frequency is equal to $\Omega_{1}=\Omega_{2}=\Omega=1.534\cdot 10^{13} \; \textrm{s}^{-\textrm{1}} $, while the two-plasmon Rabi frequency equals $\Omega^{\left(2\right)} =1.634 \cdot 10^{12} \; \textrm{s}^{-\textrm{1}} $. These values of the Rabi frequencies approximately correspond to the study~\cite{9ref}, where the NP-QD coupling factor is $1.516\cdot10^{12} \; \textrm{s}^{-\textrm{1}}$. Thus, the condition $\Omega >\Omega^{\left(2\right)} $ is fulfilled. However, we do not consider the contribution of the terms with $\Omega$ in the biexciton model (\ref{eq:12}).

It should be noted that we choose small-size QDs, while the efficiency of the biexciton formation is significantly enhanced with a larger QD~\cite{47}. On the other hand, the weak confinement regime will be satisfied for such large-size QDs and the two-level model becomes invalid. Therefore, we do not consider large QDs in this article.

Action of a magnetic field on the QD leads to a disappearance of the degeneracy of the NP-QD-NP spaser on frequency, and the possibility to simply control its frequency characteristics (see Fig.~\ref{fig:6}c). It is known that the change in the energy of an exciton in an external magnetic field depends on the orientation of the magnetic induction vector $\vec{B}_{m}$ relative to the surface of the sample~\cite{S6}. In the Faraday geometry we have a normal field orientation and the magnetic effects are most clearly manifested. There is a dual action of the magnetic field on the exciton. The action of the magnetic field occurs at the spin moment of the electron and hole, which leads to a Zeeman splitting of the exciton energy. Then the frequencies of the excitons $X_{+}$ ($X_{-}$) will take the form $\omega_{X_{-}}=\omega_{X0}-\alpha B_{m}$ ($\omega_{X_{+}}=\omega_{X0}+\alpha B_{m}$), where $\alpha=g^{F} \mu_{B}/\hbar$~\cite{S7} and $g^{F}$ is the Lande g-Factor, $\mu_{B}=9.27 \cdot 10^{-24} \; \textrm{J}/\textrm{T}$. The parameter $g^{F}$ depends on the QD radius~\cite{43}. For very small QDs, this parameter almost coincides with $g_{e}^{F} = 2$ for the free electron and decreases to $g_{CdSe}^{F} = 0.68$~\cite{gg} for the CdSe bulk semiconductor.

However, the resulting biexciton frequency $\omega_{XX0}$ and the effective detuning $\delta_{0}^{\left(2\right)}$ do not change due to Zeeman splitting~\cite{S7}. This is due to the fact that the Zeeman shifts for $X_{+}$ and $X_{-}$ compensate each other. If, however, the diamagnetic shift in the QD is taken into account, then the exciton and biexciton energies take the forms $\omega_{X_{-}}=\omega_{X0}-\alpha B_{m}+kB_{m}^{2}$ ($\omega_{X_{+}}=\omega_{X0}+\alpha B_{m}+kB_{m}^{2}$) and $\omega_{XX}=\omega_{XX0}+2kB_{m}^{2}$, where $k=\frac{e^{2}a_{ex}^{2}}{4 \hbar \mu^{*}}$. The parameters $\mu^{*}=\left(\frac{1}{m_{e}}+\frac{1}{m_{h}}\right)^{-1}$ and $a_{ex}=\sqrt{\frac{1}{2}\left(r_{e}^{2}+r_{h}^{2}\right)}$ are determined by the mass and radius of the exciton~\cite{S7}, where $r_{e}$ and $r_{h}$ are the effective radiuses of the electron and hole respectively.

Then the corresponding detunings will take the forms $\delta_{1,2}=\bar{\omega}-\omega_{X_{-},X_{+}}$ and $\delta^{\left(2\right)}=2\bar{\omega}-\omega_{XX0}-2kB_{m}^{2}$, where $\bar{\omega}$ determines the new spaser frequency in the system, taking into account the magnetic field. Expression for $\bar{\omega}$ can be obtained from the condition (\ref{eq:16}):
\begin{equation}
\label{eq:17}
\bar{\omega} =\frac{\tau_{c}\omega_{p}+\tau_{S}\omega_{XX}}{\tau_{c}+2\tau_{S}}.
\end{equation}

The last stage of parameter optimization consists of choosing a specific value of magnetic field magnitude $B_{m}$ in order to obtain the maximum entanglement between plasmons $\hat{c}_{1}$ and $\hat{c}_{2}$. This optimization results in a magnitude of magnetic field equals $B_{m}=5 \; \textrm{T}$. Since the Lande g-factor is $g^{F}_{CdSe}=1.71$ for the CdSe QD~\cite{gg} with a given size, the diamagnetic shift takes the value $kB^{2}_{m}=9.22\cdot10^{10} \; \textrm{s}^{-\textrm{1}}$ and the Zeeman shift takes the value $\alpha B_{m}=\pm 7.51 \cdot 10^{11} \; \textrm{s}^{-\textrm{1}}$. Taking into account the solution (\ref{eq:17}), the spaser frequency becomes equal to the value $\bar{\omega} =3.625 \cdot 10^{15} \; \textrm{s}^{-\textrm{1}} $ ($\delta_{1} =-1.433 \cdot 10^{14} \; \textrm{s}^{-\textrm{1}} $ and $\delta_{2} =-1.448 \cdot 10^{14} \; \textrm{s}^{-\textrm{1}} $, $\Delta=8.678 \cdot 10^{10} \; \textrm{s}^{-\textrm{1}}$, $\delta^{\left(2\right)}=-1.085 \cdot 10^{10} \; \textrm{s}^{-\textrm{1}}$). The frequency detunings $\delta_{1}$ ($\delta_{2}$) satisfy to inequality $\left|\delta_{1,2}\right|>\Omega,\frac{1}{\tau_{c}},\frac{1}{\tau_{S}}$ for the nonlinear regime in the spaser. Then the corresponding Rabi frequencies will take the values $\Omega_{1}^{\left(2\right)}=1.642 \cdot 10^{12} \; \textrm{s}^{-\textrm{1}}$, $\Omega_{2}^{\left(2\right)}=1.625 \cdot 10^{12} \; \textrm{s}^{-\textrm{1}}$.

The key point is that these Rabi frequencies can be controlled by changing the value $B_{m}$ of the external magnetic field. In particular, the detuning correction $\Delta\delta=\delta_{1}-\delta_{0}$ due to the magnetic field is only $7.456\cdot10^{11} \; \textrm{s}^{-\textrm{1}}$, which cannot influence the development of linear processes in the system with their substantially higher values of the Rabi frequencies. However, the values of $\Delta\delta$ are of the same order with $\Omega^{\left(2\right)}$ and an influence of magnetic effects on the dynamics of nonlinear interactions is possible. From a technical point of view, localization of the magnetic field on nanoscales can be achieved with the help of a magnetic cantilever.

\begin{figure}[t]
\includegraphics[width=\columnwidth]{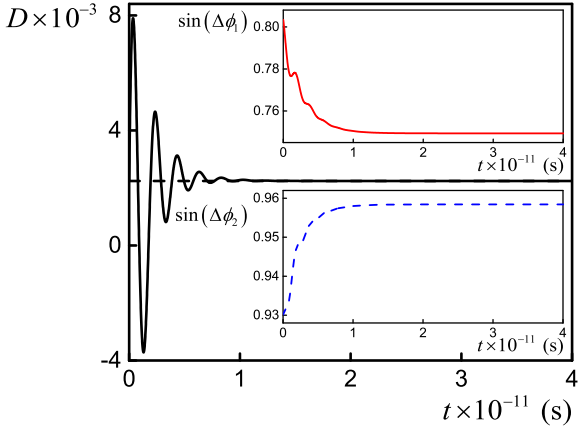}
\caption{\label{fig:7} (Color online) The time dependence of population imbalance parameter $D $ (black lines), calculated by the using formula (\ref{eq:24}) (dashed line) and by using direct numerical simulation (solid line) of the system (\ref{eq:13}). The initial parameters correspond to $D\left(0\right) =1.005\bar{D} $, $S\left(0\right) =1.005\bar{S} $ and $c_{1}\left(0\right) =0.8\bar{c}_{1} $, $c_{2}\left(0\right) =1.2\bar{c}_{2} $, for stationary values $\bar{c}_{1} =0.7084+0.7084i$, $\bar{c}_{2} =0.5009+0.8676i$, $\bar{S} =-0.1327+0.0197i$, $\bar{D}=0.0022$. Insets: dependence of the relative phase between the plasmon and exciton modes. The simulation parameters correspond to point A from the inset in Fig.~\ref{fig:8}.}
\end{figure}

\begin{figure}[t]
\includegraphics[width=\columnwidth]{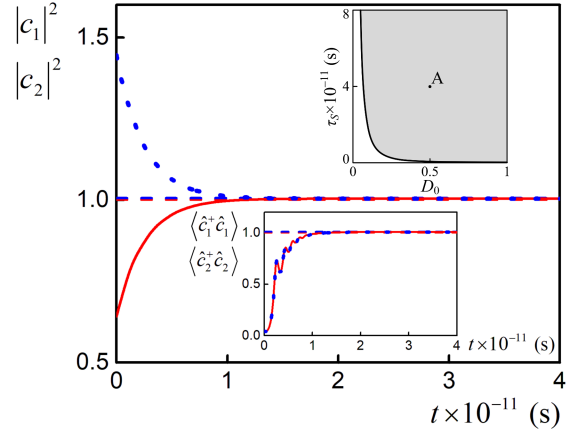}
\caption{\label{fig:8} (Color online) The dependence of the average number of plasmons $\left|\bar{c}_{1} \right|^{2} $ (solid and dashed red lines) and $\left|\bar{c}_{2} \right|^{2} $ (dotted and dashed blue lines) on time, calculated by using the formulas (\ref{eq:19}) and (\ref{eq:24}) (dashed lines) and by using direct numerical simulation (solid and dotted lines) of the system (\ref{eq:13}). Parameters of nonlinear interaction: $\omega_{p}=3.625\cdot 10^{15} \; \textrm{s}^{-\textrm{1}}$, $\Omega^{\left(2\right)}=1.625\cdot 10^{12} \; \textrm{s}^{-\textrm{1}}$, $\tau_{c}=5\cdot 10^{-12} \; \textrm{s}$ and $\tau_{S}=4\cdot 10^{-11} \; \textrm{s}$. Top inset: the stability area as a result of optimization of the nonlinear spaser with point A ($0.5$, $4\cdot 10^{-11} \; \textrm{s}$). Bottom inset: time dependence of plasmon average values for $c_{1}$ (solid and dashed red lines) and $c_{2}$ (dotted and dashed blue lines) calculated by using formulas (\ref{eq:19}) and (\ref{eq:24}) (dashed lines) and by using direct numerical simulation (solid and dotted lines) of the system (\ref{eq:B}).}
\end{figure}

The figures~\ref{fig:7} and~\ref{fig:8} show a comparison of the analytical solutions (\ref{eq:19}) and (\ref{eq:24}) and the direct numerical simulations of (\ref{eq:13}). The deviations of initial values of simulation parameters $c_{1,\left(2\right)}$ from its steady-state solutions are $20\%$, but for $S$ and $D$ these deviations are $0.5\%$. The simulation parameters were selected as follows: $D_{0} =0.5$ and $\tau_{D} =6.2\cdot 10^{-13} \; \textrm{s}$. The characteristic pumping rate $1/\tau_{D}=10^{13}$ approximately corresponds to the work~\cite{9ref}, where it takes a value of about $7.8\cdot10^{12} \; \textrm{s}^{-\textrm{1}}$.

The direct numerical simulation demonstrates the achievement of the steady-state solution for population imbalance $\bar{D} =0.0022$ within the characteristic time of $12 \; \textrm{ps}$ in Fig.~\ref{fig:7}. During the process of spaser parameters stabilization, the average number of plasmons reaches a value equalling $1$ for each NP, in Fig.~\ref{fig:8}. The spaser kinematic demonstrates the nonlinear damping oscillations of the relative phase parameter $\sin \left(\Delta \phi_{i} \left(t\right)\right)$ between the plasmon and exciton modes, where $\Delta \phi_{i} \left(t\right)=\textrm{Arg}\left(\frac{S \left(t\right)}{c_{i} \left(t\right)} \right)$ (see insets in Fig.~\ref{fig:7}).

\begin{figure}[t]
\includegraphics[width=\columnwidth]{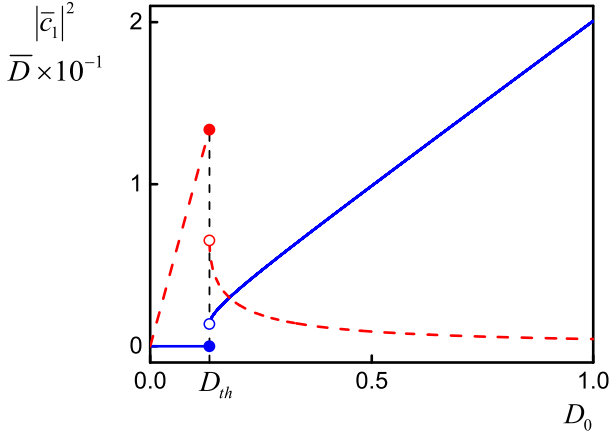}
\caption{\label{fig:9} (Color online) The gain curves for numbers of plasmons $\left|\bar{c}_{1} \right|^{2} $ (solid blue lines) and population imbalance $\bar{D} $ (dashed red lines), versus pumping value $D_{0} $. The interaction parameters correspond to point A from the top inset in Fig.~\ref{fig:8}.}
\end{figure}

The gain curves for the plasmon number in Fig.~\ref{fig:9} demonstrate a pronounced nonlinear behavior. At the initial stage, the pumping action leads to a linear increase of the population imbalance only, while the average number of plasmons $\left|\bar{c}_{i}\right|^{2}$ does not change and is zero. When the pump achieves the threshold value $D_{0}=D_{th}=0.1337$ first-order discontinuity appears with a jump in the population imbalance and the average number of plasmons. A further increase in the pump value leads to an increase in the number of plasmons with a simultaneous decrease in the population imbalance parameter. This effect cannot be observed in the linear case shown in Fig.~\ref{fig:2}, because this feature is due to a fundamental dependence of the stationary solution $\bar{D} $ on the value of the external pump $D_{0} $ in (\ref{eq:24}). In particular, after reaching the generation threshold in a linear system, the pump energy is distributed between the generation of plasmons and excitons in approximately equally amounts (see~\cite{7} and the inset in Fig.~\ref{fig:2}). Thus, the advantage of the nonlinear spaser model is the more efficient transfer of pump energy to the generation of plasmon modes.

Now, we will analyze the dynamics of the parameter
\begin{equation}
\label{eq:25}
G_{12}^{\left(2\right)}\left(t,\tau\right) = \frac{\left\langle \hat{c}_{1}^{+}\left(t\right) \hat{c}_{1}\left(t\right) \hat{c}_{2}^{+}\left(t+\tau\right) \hat{c}_{2}\left(t+\tau\right) \right\rangle}{\left\langle \hat{c}_{1}^{+}\left(t\right) \hat{c}_{1}\left(t\right) \right\rangle \left\langle \hat{c}_{2}^{+}\left(t+\tau\right) \hat{c}_{2}\left(t+\tau\right)\right \rangle},
\end{equation}
which corresponds to the cross-correlation function and is a criterion for establishing correlations between plasmonic modes $\hat{c}_{1} $ and $\hat{c}_{2} $. In particular, the condition $G_{12}^{\left(2\right)} \equiv G_{12}^{\left(2\right)}\left(t,0\right) > 1$ is associated with the intermode plasmon bunching. Moreover, the violation of the Cauchy-Schwarz inequality $$C \equiv \frac{\left.G_{12}^{\left(2\right)}\left(t,0\right)\right.^{2}}{g_{1}^{\left(2\right)}\left(t,0\right) g_{2}^{\left(2\right)}\left(t,0\right)} \leq 1$$ indicates the nonclassical character of the correlations between plasmonic modes, where $g_{i}^{\left(2\right)}\left(t,0\right)=G_{i}^{\left(2\right)}\left(t,0\right)+1$, $i=1,2$. This is the necessary condition for the generation of entangled plasmons in the spaser system.

\begin{figure}[t]
\includegraphics[width=\columnwidth]{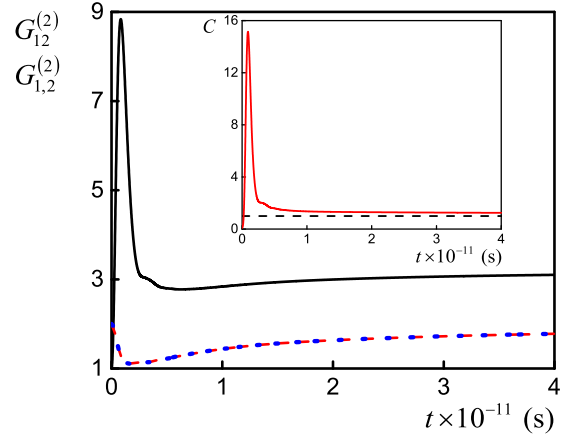}
\caption{\label{fig:10} (Color online) The time dependence of cross-correlation function $G_{12}^{\left(2\right)}$ (solid black line) and autocorrelation functions $G_{1}^{\left(2\right)}$ (dashed red line) and $G_{2}^{\left(2\right)}$ (dotted blue line) for plasmons in nonlinear NP-QD-NP spaser. Inset: time dependence of $C$ parameter (solid red line) and the unity value of this parameter $C=1$ (dashed black line) indicating on the violation of Cauchy-Schwarz inequality for generated plasmons in region above it.}
\end{figure}

In this part of the article we aim to show the formation of strong nonclassical correlations between two plasmons, which are generated during the biexciton decay process in the NP-QD-NP spaser system. In order to demonstrate this effect, we solve the system of kinematic equations (\ref{eq:B}) for the correlators of plasmon and exciton modes, which was derived on the basis of system (\ref{eq:13}). In particular, the numerical simulation of system (\ref{eq:B}) allows us to obtain the time dependence of the cross-correlation function $G_{12}^{\left(2\right)} $ (see Fig.~\ref{fig:10}). Fig.~\ref{fig:10} shows the result of such simulation with system parameters as for Figs.~\ref{fig:7},~\ref{fig:8}.

The initial values of intramode correlators were chosen as follows:
\begin{eqnarray}
\nonumber
&\left.\left\langle\hat{S}^{+}\hat{S}\right\rangle\right|_{t=0}=\left|S\left(0\right)\right|^{2},& \\
\nonumber
&\left.\left\langle\hat{S}\hat{S}^{+}\right\rangle\right|_{t=0}=\left|S\left(0\right)\right|^{2}-D\left(0\right),& \\
\nonumber
&\left.\left\langle \hat{S}^{2} \right\rangle\right|_{t=0}=S^{2}\left(0\right), \; \left.\left\langle \left(\hat{S}^{+}\right)^{2} \right\rangle\right|_{t=0}=\left(S^{*}\left(0\right)\right)^{2};& \\
\nonumber
&\left.\left\langle\hat{c}_{i}^{+}\hat{c}_{i}\right\rangle\right|_{t=0}=\left|c_{i}\left(0\right)\right|^{2}, \; \left.\left\langle\hat{c}_{i}\hat{c}_{i}^{+}\right\rangle\right|_{t=0}=1+\left|c_{i}\left(0\right)\right|^{2},& \\
\nonumber
&\left.\left\langle \hat{c}_{i}^{2} \right\rangle\right|_{t=0}=c_{i}^{2}\left(0\right), \; \left.\left\langle \left(\hat{c}_{i}^{+}\right)^{2} \right\rangle\right|_{t=0}=\left(c_{i}^{*}\left(0\right)\right)^{2},&
\end{eqnarray}
where $i\in\left\{1,2\right\}$. However, the initial values for intermode plasmonic and plasmon-exciton correlators are assumed to equal zero, i.e.
\begin{eqnarray}
\nonumber
&\left.\left\langle\hat{c}_{i}^{+}\hat{c}_{j}\right\rangle\right|_{t=0}=
\left.\left\langle\hat{c}_{i}\hat{c}_{j}^{+}\right\rangle\right|_{t=0}=
\left.\left\langle\hat{c}_{i}\hat{S}\right\rangle\right|_{t=0}=
\left.\left\langle\hat{S}\hat{c}_{i}\right\rangle\right|_{t=0}=&\\
\nonumber
&\left.\left\langle\hat{c}_{i}^{+}\hat{S}\right\rangle\right|_{t=0}=
\left.\left\langle\hat{S}\hat{c}_{i}^{+}\right\rangle\right|_{t=0}=
\left.\left\langle\hat{c}_{i}\hat{S}^{+}\right\rangle\right|_{t=0}=&\\
\nonumber
&\left.\left\langle\hat{S}^{+}\hat{c}_{i}\right\rangle\right|_{t=0}=
\left.\left\langle\hat{c}_{i}^{+}\hat{S}^{+}\right\rangle\right|_{t=0}=
\left.\left\langle\hat{S}^{+}\hat{c}_{i}^{+}\right\rangle\right|_{t=0}=0,&
\end{eqnarray}
where $i,j\in\left\{1,2\right\}$ and $i \neq j$.
During the process of nonlinear interaction between QD and NPs in the spaser, the correlators $\left\langle\hat{c}_{i}^{+}\hat{c}_{i}\right\rangle$ are stabilized at values equal to $1$ in full agreement with mean-field theory (see Fig.~\ref{fig:8}). At the same time, the dynamics of the establishment of a stationary regime based on simulation of the system (\ref{eq:B}) (see bottom inset in Fig.~\ref{fig:8}) differ from the dynamics based on simulation of the system for average values (\ref{eq:13}) (see Fig.~\ref{fig:8}). In particular,
the appearance of the plasmon correlations during the time evolution of the system (\ref{eq:B}) leads to a more oscillatory establishment of the stationary regime (see bottom inset in Fig.~\ref{fig:8}).

The value of parameter $G_{12}^{\left(2\right)} $ in such stationary conditions reaches the level $3.15$, which demonstrates strong intermode bunching between modes $\hat{c}_{1} $ and $\hat{c}_{2} $. Moreover, the maximum value of $C$ parameter is $C_{\textrm{max}}=15.15$ and the stationary regime value is $C_{\textrm{st}}=1.2$, which demonstrates the nonclassical character of correlations between plasmon modes in spaser system, see inset in Fig.~\ref{fig:10}. This corresponds to the case when the initial entanglement between plasmons $\hat{c}_{1} $ and $\hat{c}_{2} $ is completely absent. We also analyzed the autocorrelation functions $G_{1}^{\left(2\right)}$ for $\hat{c}_{1}$ and $G_{2}^{\left(2\right)}$ for $\hat{c}_{2}$ plasmon fields in accordance with (\ref{eq:11}). The values of autocorrelation functions achieve the level $1.87$, which corresponds to the super-Poisson statistics (see Fig.~\ref{fig:10}).

Thus, the main result of our simulation is a demonstration of the development of quantum correlations between two localized plasmonic modes and the possibility for formation of an entangled state during the process of biexciton state decay in a nonlinear NP-QD-NP spaser system. We also note that if the nonlinear biexciton$\rightarrow$plasmons energy exchange (\ref{eq:12}) is replaced by linear plasmon-exciton interaction of type (\ref{eq:4}), the formation of quantum correlations in the pair of plasmons does not occur.

A simple way to influence the quantum properties of the generated nonclassical plasmons in a spaser is to change the value of the magnetic field. Thus, by increasing the field induction $B_{m}$, the Zeeman splitting and diamagnetic shift of the QD levels increase, the parameters $\delta_{1,2}$, $\delta^{\left(2\right)}$ and $\Omega^{\left(2\right)}$ change, and the correlation functions in Fig.~\ref{fig:10} achieve a new level of values.

It should be noted that the loss of stability of plasmon generation regimes occurs, in general, due to chemical impurity of the individual components -- NPs and QDs, and also the inaccuracy of their position in the spaser. In addition, it is necessary to take into account that the efficiency of the magnetic field's influence on the quantum state of the excitons also depends on the shape of the QD~\cite{48}.

Another important problem is the temporal stability of QD parameters during the establishment process of the stationary regime of the spaser, similar to the problem of "blinking" for emitting QDs~\cite{49}. One approach to solving this problem is the use of a composite core-shell NP~\cite{50}.

\section{Conclusion}
We have investigated the dynamics of the average number of localized plasmons and their quantum statistics in the double spaser system. The system consists of two NPs and two QDs located in the vertices of a square and coupled with each other by means of a near-field interaction. Based on the realistic parameters of gold NPs and CdSe semiconductor QDs, we have optimized the geometric and dissipative characteristics of the spaser system for the observation of steady-state solutions for average numbers of plasmons and excitons in the system. We have studied the features of the quantum statistics of the generated plasmons in the spaser. In particular, it has been shown that due to the dipole-dipole interactions between adjacent metallic NPs in the $2 \times 2$ spaser, a significant decrease of the autocorrelation function value $G_{1}^{\left(2\right)} \left(t\right)$ for plasmons can be observed. Meanwhile, in the absence of these dipole-dipole NP interactions in the $1 \times 1$ spaser, the value $G_{1}^{\left(2\right)} \left(t\right)$ remains at the level of $2$ with pronounced super-Poisson statistics. Using this fact, we assumed that systems of linear spasers are not suitable for the formation and control of nonclassical states of plasmons. At the same time, our result supplements the model in~\cite{16a}, in which strong NP-NP dipole-dipole interactions for the spaser-like system have been investigated with respect to the nonclassicality of \textit{photon} states.

We have also proposed a novel mechanism for the control of plasmons' quantum properties in the spaser, where a pair of NPs is coupled by nonlinear near-field interactions with a single QD (NP-QD-NP spaser). We have optimized the size of the QD and the geometry of the spaser system to increase the efficiency of far-off-resonant plasmon-exciton interaction between the QD and NPs. As a result, optimal conditions for two-quantum biexciton decay with the appearance of plasmons, which are localized on the corresponding NPs, have been formulated. We have introduced an additional degree of freedom in the form of an external magnetic field acting in our spaser system. It has been shown that the influence of an external magnetic field on such a system firstly leads to the Zeeman splitting and secondly diamagnetic shift of QD energy sublevels, and changes in the two-plasmon Rabi frequency and plasmon-exciton detuning in the spaser.

Finally, we have optimized the system parameters providing the stationary regime of the entangled plasmons generation. We have proposed to use this regime for the experimental generation of biplasmons and their further use in quantum plasmonic circuits. In particular, the integration of the spaser system into the plasmonic chip can allow the creation of a source for generating correlated states of the electromagnetic field, it could be important for application in quantum computing at the nanoscale. Further development of this work may be aimed on a complex simulation of quantum algorithms (including soliton-like structures~\cite{50a}) in a chain of nonlinear spasers, taking into account the nonlinear magnetic effects~\cite{51}, multipole resonances~\cite{5ref,Evlyukhin} and collective effects as in optics~\cite{52}.

\begin{acknowledgments}
We would like to thank A.B. Evlyukhin for helpful discussions. This work was supported by the Russian Foundation for Basic Research Grant No. 17-42-330001 and the Ministry of Education and Science of the Russian Federation within the state task VlSU 2017 in the field of scientific research and the Michail-Lomonosov-Programme - Linie~A,~2017 (ID~57320203) of the German Academic Exchange Service and the Ministry of Education and Science of the Russian Federation (No.~2.9953.2017/DAAD).
\end{acknowledgments}

\appendix

\section{\label{app:A}System of equations on average values for bi-linear combinations for calculation of autocorrelation function $G_{i}^{\left(2\right)} \left(t\right)$}
For analysis of the dynamics of parameter $G_{i}^{\left(2\right)} \left(t\right)$, it is convenient to use Wick's theorem for decreasing the degrees of correlators in (\ref{eq:11}) and using (\ref{eq:6}) to reduce the problem to a system of equations on average values for bi-linear combinations:
{\allowdisplaybreaks
\begin{widetext}
\begin{subequations}
\label{eq:A}
\begin{eqnarray}
\nonumber
\frac{d}{dt} \left\langle (\hat{c}_{1}^{+})^{2} (\hat{c}_{1})^{2} \right\rangle &=&-\frac{4}{\tau_{c} } \left\langle (\hat{c}_{1}^{+})^{2} (\hat{c}_{1})^{2} \right\rangle +2i\Omega \left(\left\langle \hat{c}_{1}^{2} \right\rangle \left\langle \hat{c}_{1}^{+} \hat{S}_{1}^{+} \right\rangle -\left\langle (\hat{c}_{1}^{+})^{2} \right\rangle \left\langle \hat{c}_{1} \hat{S}_{1} \right\rangle +2\left\langle \hat{c}_{1}^{+} \hat{c}_{1} \right\rangle \left(\left\langle \hat{S}_{1}^{+} \hat{c}_{1} \right\rangle -\left\langle \hat{c}_{1}^{+} \hat{S}_{1} \right\rangle \right)\right) \\
&& +2i\Omega_{pp} \left(\left\langle \hat{c}_{1}^{2} \right\rangle \left\langle \hat{c}_{1}^{+} \hat{c}_{2}^{+} \right\rangle -\left\langle (\hat{c}_{1}^{+})^{2} \right\rangle \left\langle \hat{c}_{1} \hat{c}_{2} \right\rangle +2\left\langle \hat{c}_{1}^{+} \hat{c}_{1} \right\rangle \left(\left\langle \hat{c}_{2}^{+} \hat{c}_{1} \right\rangle -\left\langle \hat{c}_{1}^{+} \hat{c}_{2} \right\rangle \right)\right), \\
\frac{d}{dt} \left\langle \hat{c}_{1,2}^{2} \right\rangle &=&2iA\left\langle \hat{c}_{1,2}^{2} \right\rangle -2i\Omega \left\langle \hat{c}_{1,2} \hat{S}_{1,2} \right\rangle -2i\Omega _{pp} \left\langle \hat{c}_{1} \hat{c}_{2} \right\rangle, \\
\nonumber
\frac{d}{dt} \left\langle \hat{c}_{1,2} \hat{c}_{1,2}^{+} \right\rangle &=&-\frac{2}{\tau_{c} } \left\langle \hat{c}_{1,2} \hat{c}_{1,2}^{+} \right\rangle -i\Omega \left(\left\langle \hat{S}_{1,2} \hat{c}_{1,2}^{+} \right\rangle -\left\langle \hat{c}_{1,2} \hat{S}_{1,2}^{+} \right\rangle \right) \\
&& -i\Omega_{pp} \left(\left\langle \hat{c}_{2,1} \hat{c}_{1,2}^{+} \right\rangle -\left\langle \hat{c}_{1,2} \hat{c}_{2,1}^{+} \right\rangle \right)+\left\langle \hat{F}_{c1,c2} \hat{c}_{1,2}^{+} \right\rangle +\left\langle \hat{c}_{1,2} \hat{F}_{c1,c2}^{+} \right\rangle, \\
\frac{d}{dt} \left\langle \hat{c}_{1,2}^{+} \hat{c}_{1,2} \right\rangle &=&-\frac{2}{\tau_{c} } \left\langle \hat{c}_{1,2}^{+} \hat{c}_{1,2} \right\rangle +i\Omega \left(\left\langle \hat{S}_{1,2}^{+} \hat{c}_{1,2} \right\rangle -\left\langle \hat{c}_{1,2}^{+} \hat{S}_{1,2} \right\rangle \right)+i\Omega_{pp} \left(\left\langle \hat{c}_{2,1}^{+} \hat{c}_{1,2} \right\rangle -\left\langle \hat{c}_{1,2}^{+} \hat{c}_{2,1} \right\rangle \right), \\
\frac{d}{dt} \left\langle \hat{c}_{1,2} \hat{S}_{1,2} \right\rangle &=&i\left(A+B\right)\left\langle \hat{c}_{1,2} \hat{S}_{1,2} \right\rangle +i\Omega D_{1,2} \left\langle \hat{c}_{1,2}^{2} \right\rangle -i\Omega \left\langle \hat{S}_{1,2}^{2} \right\rangle, \\
\frac{d}{dt} \left\langle \hat{S}_{1,2}^{+} \hat{c}_{1,2} \right\rangle &=&i\left(A-B^{*} \right)\left\langle \hat{S}_{1,2}^{+} \hat{c}_{1,2} \right\rangle -i\Omega \left(\left\langle \hat{S}_{1,2}^{+} \hat{S}_{1,2} \right\rangle +D_{1,2} \left\langle \hat{c}_{1,2}^{+} \hat{c}_{1,2} \right\rangle \right), \\
\frac{d}{dt} \left\langle \hat{c}_{1} \hat{c}_{2} \right\rangle &=&2iA\left\langle \hat{c}_{1} \hat{c}_{2} \right\rangle -i\Omega_{pp} \left(\left\langle \hat{c}_{1}^{2} \right\rangle +\left\langle \hat{c}_{2}^{2} \right\rangle \right), \\
\frac{d}{dt} \left\langle \hat{c}_{2}^{+} \hat{c}_{1} \right\rangle &=&-\frac{2}{\tau_{c} } \left\langle \hat{c}_{2}^{+} \hat{c}_{1} \right\rangle -i\Omega_{pp} \left(\left\langle \hat{c}_{2}^{+} \hat{c}_{2} \right\rangle -\left\langle \hat{c}_{1}^{+} \hat{c}_{1} \right\rangle \right), \\
\frac{d}{dt} \left\langle \hat{S}_{1,2} \hat{c}_{1,2}^{+} \right\rangle &=&i\left(B-A^{*} \right)\left\langle \hat{S}_{1,2} \hat{c}_{1,2}^{+} \right\rangle +i\Omega \left(D_{1,2} \left\langle \hat{c}_{1,2} \hat{c}_{1,2}^{+} \right\rangle +\left\langle \hat{S}_{1,2} \hat{S}_{1,2}^{+} \right\rangle \right), \\
\frac{d}{dt} \left\langle \hat{c}_{2} \hat{c}_{1}^{+} \right\rangle &=&-\frac{2}{\tau_{c} } \left\langle \hat{c}_{2} \hat{c}_{1}^{+} \right\rangle -i\Omega_{pp} \left(\left\langle \hat{c}_{1} \hat{c}_{1}^{+} \right\rangle -\left\langle \hat{c}_{2} \hat{c}_{2}^{+} \right\rangle \right), \\
\frac{d}{dt} \left\langle \hat{S}_{1,2}^{2} \right\rangle &=&2iB\left\langle \hat{S}_{1,2}^{2} \right\rangle +2i\Omega D_{1,2} \left\langle \hat{c}_{1,2} \hat{S}_{1,2} \right\rangle, \\
\frac{d}{dt} \left\langle \hat{S}_{1,2} \hat{S}_{1,2}^{+} \right\rangle &=&-\frac{2}{\tau_{S} } \left\langle \hat{S}_{1,2} \hat{S}_{1,2}^{+} \right\rangle +i\Omega D_{1,2} \left( \left\langle \hat{c}_{1,2} \hat{S}_{1,2}^{+} \right\rangle - \left\langle \hat{S}_{1,2} \hat{c}_{1,2}^{+} \right\rangle \right)+\left\langle \hat{F}_{S1,S2} \hat{S}^{+}_{1,2} \right\rangle+\left\langle \hat{S}_{1,2} \hat{F}^{+}_{S1,S2} \right\rangle, \\
\frac{d}{dt} \left\langle \hat{S}_{1,2}^{+} \hat{S}_{1,2} \right\rangle &=&-\frac{2}{\tau_{S} } \left\langle \hat{S}_{1,2}^{+} \hat{S}_{1,2} \right\rangle -i\Omega D_{1,2} \left( \left\langle \hat{c}_{1,2}^{+} \hat{S}_{1,2} \right\rangle - \left\langle \hat{S}_{1,2}^{+} \hat{c}_{1,2} \right\rangle \right), \\
\frac{d}{dt} D_{1,2} &=&-2i\Omega \left(\left\langle \hat{S}_{1,2}^{+} \hat{c}_{1,2} \right\rangle -\left\langle \hat{S}_{1,2} \hat{c}_{1,2}^{+} \right\rangle \right)-\frac{D_{1,2} -D_{0} }{\tau_{D} },
\end{eqnarray}
\end{subequations}
\end{widetext}
}
\noindent where $$A=\Delta +\frac{i}{\tau_{c} }, \; B=\delta +\frac{i}{\tau_{S} }, \; D_{1,2} \equiv \left\langle D_{1,2} \right\rangle,$$ the average values of noise correlation functions have the form
$$\left\langle \hat{F}_{c1,c2} \hat{c}_{1,2}^{+} \right\rangle =\left\langle \hat{c}_{1,2} \hat{F}_{c1,c2}^{+} \right\rangle =\frac{Z_{c1,c2} }{2},$$
$$\left\langle \hat{F}_{S1,S2} \hat{S}_{1,2}^{+} \right\rangle =\left\langle \hat{S}_{1,2} \hat{F}_{S1,S2}^{+} \right\rangle =\frac{Z_{S1,S2} }{2},$$
$$Z_{c1,c2} =\frac{2}{\tau_{c} }, \; Z_{S1,S2} =-\frac{2}{\tau_{S} } D_{1,2} -\dot{D}_{1,2}$$
with consideration for commutation relationships for $\hat{c}_{i} $ and $\hat{S}_{i} $~\cite{Scully}.

~\

~\

~\

~\

~\

~\

~\

\section{\label{app:B}System of equations on average values for bi-linear combinations for calculation of cross-correlation function $G_{12}^{\left(2\right)} \left(t\right)$}
In the conditions of the current nonlinear problem (\ref{eq:13}) for the visualization of function (\ref{eq:25}), it is necessary to decouple correlators and decrease their degrees with the help of Wick's theorem. As a result, the complete self-consistent system on correlators takes the form:
{\allowdisplaybreaks
\begin{widetext}
\begin{subequations}
\label{eq:B}
\begin{eqnarray}
\frac{d}{dt} \left\langle \hat{c}_{1,2} \right\rangle &=&A_{1}\left\langle \hat{c}_{1,2} \right\rangle -i\Omega ^{\left(2\right)} \left\langle \hat{c}_{2,1}^{+} S\right\rangle, \\
\frac{d}{dt} \left\langle \hat{S}\right\rangle &=&B_{1}\left\langle \hat{S}\right\rangle +i\Omega ^{\left(2\right)} \left\langle \hat{c}_{1} \hat{c}_{2} \right\rangle D, \\
\frac{d}{dt} \left\langle \hat{c}_{1,2}^{2} \right\rangle &=&2A_{1}\left\langle \hat{c}_{1,2}^{2} \right\rangle -2i\Omega ^{\left(2\right)} \left(\left\langle \hat{c}_{2,1}^{+} \hat{c}_{1,2} \right\rangle \left\langle \hat{S}\right\rangle +\left\langle \hat{c}_{1,2} \hat{S}\right\rangle \left\langle \hat{c}_{2,1}^{+} \right\rangle +\left\langle \hat{c}_{2,1}^{+} \hat{S}\right\rangle \left\langle \hat{c}_{1,2} \right\rangle \right), \\
\frac{d}{dt} \left\langle \hat{S}^{2} \right\rangle &=&2B_{1}\left\langle \hat{S}^{2} \right\rangle +2i\Omega ^{\left(2\right)} D\left(\left\langle \hat{c}_{1} \hat{c}_{2} \right\rangle \left\langle \hat{S}\right\rangle +\left\langle \hat{c}_{1} \hat{S}\right\rangle \left\langle \hat{c}_{2} \right\rangle +\left\langle \hat{c}_{2} \hat{S}\right\rangle \left\langle \hat{c}_{1} \right\rangle \right), \\
\nonumber
\frac{d}{dt} \left\langle \hat{c}_{1,2}^{+} \hat{c}_{1,2} \right\rangle &=&-\frac{2}{\tau _{C} } \left\langle \hat{c}_{1,2}^{+} \hat{c}_{1,2} \right\rangle +i\Omega ^{\left(2\right)} \Bigl(\left\langle \hat{c}_{1} \hat{c}_{2} \right\rangle \left\langle S^{+} \right\rangle -\left\langle \hat{c}_{1}^{+} \hat{c}_{2}^{+} \right\rangle \left\langle \hat{S}\right\rangle +\left\langle \hat{c}_{1} \hat{S}^{+} \right\rangle \left\langle \hat{c}_{2} \right\rangle -\left\langle \hat{c}_{1}^{+} \hat{S}\right\rangle \left\langle \hat{c}_{2}^{+} \right\rangle \\
&& \mspace{394mu}
+\left\langle \hat{c}_{2} \hat{S}^{+} \right\rangle \left\langle \hat{c}_{1} \right\rangle -\left\langle \hat{c}_{2}^{+} \hat{S}\right\rangle \left\langle \hat{c}_{1}^{+} \right\rangle \Bigr), \\
\nonumber
\frac{d}{dt} \left\langle \hat{c}_{1,2} \hat{c}_{1,2}^{+} \right\rangle &=&-\frac{2}{\tau _{C} } \left\langle \hat{c}_{1,2} \hat{c}_{1,2}^{+} \right\rangle +i\Omega ^{\left(2\right)} \Bigl(\left\langle \hat{c}_{1} \hat{c}_{2} \right\rangle \left\langle S^{+} \right\rangle -\left\langle \hat{c}_{1}^{+} \hat{c}_{2}^{+} \right\rangle \left\langle \hat{S}\right\rangle +\left\langle \hat{c}_{1} \hat{S}^{+} \right\rangle \left\langle \hat{c}_{2} \right\rangle -\left\langle \hat{c}_{1}^{+} \hat{S}\right\rangle \left\langle \hat{c}_{2}^{+} \right\rangle \\
&& \mspace{176mu}
+\left\langle \hat{c}_{2} \hat{S}^{+} \right\rangle \left\langle \hat{c}_{1} \right\rangle -\left\langle \hat{c}_{2}^{+} \hat{S}\right\rangle \left\langle \hat{c}_{1}^{+} \right\rangle \Bigr)+\left\langle \hat{F}_{c1,c2} \hat{c}_{1,2}^{+} \right\rangle +\left\langle \hat{c}_{1,2} \hat{F}_{c1,c2}^{+} \right\rangle , \\
\nonumber
\frac{d}{dt} \left\langle \hat{S}^{+} \hat{S}\right\rangle &=&-\frac{2}{\tau _{S} } \left\langle \hat{S}^{+} \hat{S}\right\rangle +i\Omega ^{\left(2\right)} D\Bigl(\left\langle \hat{c}_{1} \hat{c}_{2} \right\rangle \left\langle S^{+} \right\rangle -\left\langle \hat{c}_{1}^{+} \hat{c}_{2}^{+} \right\rangle \left\langle \hat{S}\right\rangle +\left\langle \hat{c}_{1} \hat{S}^{+} \right\rangle \left\langle \hat{c}_{2} \right\rangle -\left\langle \hat{c}_{1}^{+} \hat{S}\right\rangle \left\langle \hat{c}_{2}^{+} \right\rangle \\
&& \mspace{394mu}
+\left\langle \hat{c}_{2} \hat{S}^{+} \right\rangle \left\langle \hat{c}_{1} \right\rangle -\left\langle \hat{c}_{2}^{+} \hat{S}\right\rangle \left\langle \hat{c}_{1}^{+} \right\rangle \Bigr), \\
\nonumber
\frac{d}{dt} \left\langle \hat{S}\hat{S}^{+} \right\rangle &=&-\frac{2}{\tau _{S} } \left\langle \hat{S}\hat{S}^{+} \right\rangle +i\Omega ^{\left(2\right)} D\Bigl(\left\langle \hat{c}_{1} \hat{c}_{2} \right\rangle \left\langle S^{+} \right\rangle -\left\langle \hat{c}_{1}^{+} \hat{c}_{2}^{+} \right\rangle \left\langle \hat{S}\right\rangle +\left\langle \hat{c}_{1} \hat{S}^{+} \right\rangle \left\langle \hat{c}_{2} \right\rangle -\left\langle \hat{c}_{1}^{+} \hat{S}\right\rangle \left\langle \hat{c}_{2}^{+} \right\rangle \\
&& \mspace{278mu}
+\left\langle \hat{c}_{2} \hat{S}^{+} \right\rangle \left\langle \hat{c}_{1} \right\rangle -\left\langle \hat{c}_{2}^{+} \hat{S}\right\rangle \left\langle \hat{c}_{1}^{+} \right\rangle \Bigr)+\left\langle \hat{F}_{S} \hat{S}^{+} \right\rangle +\left\langle \hat{S}\hat{F}_{S}^{+} \right\rangle, \\
\nonumber
\frac{d}{dt} \left\langle \hat{c}_{1} \hat{c}_{2} \right\rangle &=&2A_{1}\left\langle \hat{c}_{1} \hat{c}_{2} \right\rangle -i\Omega ^{\left(2\right)} \Bigl(\left\langle \hat{c}_{1} \hat{c}_{1}^{+} \right\rangle \left\langle \hat{S}\right\rangle +\left\langle \hat{c}_{2}^{+} \hat{c}_{2} \right\rangle \left\langle \hat{S}\right\rangle +\left\langle \hat{c}_{1}^{+} \hat{S}\right\rangle \left\langle \hat{c}_{1} \right\rangle +\left\langle \hat{c}_{2}^{+} \hat{S}\right\rangle \left\langle \hat{c}_{2} \right\rangle \\
&& \mspace{348mu}
+ \left\langle \hat{c}_{1} \hat{S}\right\rangle \left\langle \hat{c}_{1}^{+} \right\rangle +\left\langle \hat{c}_{2} \hat{S}\right\rangle \left\langle \hat{c}_{2}^{+} \right\rangle \Bigr), \\
\frac{d}{dt} \left\langle \hat{c}_{1}^{+} \hat{c}_{2} \right\rangle &=&\left(A_{1}+A_{1}^{*} \right)\left\langle \hat{c}_{1}^{+} \hat{c}_{2} \right\rangle +i\Omega ^{\left(2\right)} \left(\left\langle \hat{c}_{2}^{2} \right\rangle \left\langle \hat{S}^{+} \right\rangle +2\left\langle \hat{c}_{2} \hat{S}^{+} \right\rangle \left\langle \hat{c}_{2} \right\rangle -\left\langle \left.\hat{c}_{1}^{+} \right.^{2} \right\rangle \left\langle \hat{S}\right\rangle -2\left\langle \hat{c}_{1}^{+} \hat{S}\right\rangle \left\langle \hat{c}_{1}^{+} \right\rangle \right), \\
\nonumber
\frac{d}{dt} \left\langle \hat{c}_{1,2} \hat{S}\right\rangle &=&\left(A_{1}+B_{1}\right)\left\langle \hat{c}_{1,2} \hat{S}\right\rangle -i\Omega ^{\left(2\right)} \Bigl(2\left\langle \hat{c}_{2,1}^{+} \hat{S}\right\rangle \left\langle \hat{S}\right\rangle +\left\langle \hat{S}^{2} \right\rangle \left\langle \hat{c}_{2,1}^{+} \right\rangle \\
&& \mspace{330mu}
-D\left(\left\langle \hat{c}_{1,2}^{2} \right\rangle \left\langle \hat{c}_{2,1} \right\rangle +2\left\langle \hat{c}_{1} \hat{c}_{2} \right\rangle \left\langle \hat{c}_{1,2} \right\rangle \right)\Bigr), \\
\nonumber
\frac{d}{dt} \left\langle \hat{c}_{1,2}^{+} \hat{S}\right\rangle &=&\left(A_{1}^{*} +B_{1}\right)\left\langle \hat{c}_{1,2}^{+} \hat{S}\right\rangle +i\Omega ^{\left(2\right)} \Bigl(\left\langle \hat{c}_{2,1} \hat{S}^{+} \right\rangle \left\langle \hat{S}\right\rangle +\left\langle \hat{c}_{2,1} \hat{S}\right\rangle \left\langle \hat{S}^{+} \right\rangle +\left\langle \hat{S}^{+} \hat{S}\right\rangle \left\langle \hat{c}_{2,1} \right\rangle \\
&& \mspace{200mu}
+D\left(\left\langle \hat{c}_{1,2}^{+} \hat{c}_{1,2} \right\rangle \left\langle \hat{c}_{2,1} \right\rangle +\left\langle \hat{c}_{1,2}^{+} \hat{c}_{2,1} \right\rangle \left\langle \hat{c}_{1,2} \right\rangle +\left\langle \hat{c}_{1} \hat{c}_{2} \right\rangle \left\langle \hat{c}_{1,2}^{+} \right\rangle \right)\Bigr), \\
\nonumber
\frac{d}{dt} D&=&2i\Omega ^{\left(2\right)} \Bigl(\left\langle \hat{c}_{1}^{+} \hat{c}_{2}^{+} \right\rangle \left\langle \hat{S}\right\rangle -\left\langle \hat{c}_{1} \hat{c}_{2} \right\rangle \left\langle S^{+} \right\rangle +\left\langle \hat{c}_{1}^{+} \hat{S}\right\rangle \left\langle \hat{c}_{2}^{+} \right\rangle -\left\langle \hat{c}_{1} \hat{S}^{+} \right\rangle \left\langle \hat{c}_{2} \right\rangle \\
&& \mspace{268mu}
+\left\langle \hat{c}_{2}^{+} \hat{S}\right\rangle \left\langle \hat{c}_{1}^{+} \right\rangle -\left\langle \hat{c}_{2} \hat{S}^{+} \right\rangle \left\langle \hat{c}_{1} \right\rangle \Bigr)-\frac{D-D_{0} }{\tau _{D} },
\end{eqnarray}
\end{subequations}
\end{widetext}
}
\noindent where $$A_{1} =i\Delta -\frac{1}{\tau_{c} }, \; B_{1} =i\delta^{\left(2\right)} -\frac{1}{\tau_{S} }, \; D \equiv \left\langle D \right\rangle,$$
$$\delta^{\left(2\right)}=2\bar{\omega}-\omega_{XX}, \; \Delta=\bar{\omega}-\omega_{p},$$
$$\bar{\omega} =\frac{\tau_{c}\omega_{p}+\tau_{S}\omega_{XX}}{\tau_{c}+2\tau_{S}},$$
the average values of noise correlation functions have the form
\begin{eqnarray}
\nonumber
&\left\langle \hat{F}_{c1,c2} \hat{c}_{1,2}^{+} \right\rangle =\left\langle \hat{c}_{1,2} \hat{F}_{c1,c2}^{+} \right\rangle =\frac{Z_{c1,c2} }{2},& \\
\nonumber
&\left\langle \hat{F}_{S} \hat{S}^{+} \right\rangle =\left\langle \hat{S} \hat{F}_{S}^{+} \right\rangle =\frac{Z_{S} }{2},& \\
\nonumber
&Z_{c1,c2} =\frac{2}{\tau_{c} }, \; Z_{S} =-\frac{2}{\tau_{S} } D -\dot{D}&
\end{eqnarray}
with consideration for commutation relationships for $\hat{c}_{i} $ and $\hat{S} $~\cite{Scully}.

\bibliography{Main_text_incl._figures}

\end{document}